\documentclass[11pt,a4paper]{atlasnote}
\pdfoutput=1
\usepackage{epstopdf}
\usepackage{graphicx}
\usepackage{caption}
\usepackage[labelformat=simple]{subcaption}
\usepackage{defphysics}
\usepackage{amsmath}
\usepackage{amssymb}
\usepackage{mathptmx}
\usepackage{booktabs}
\usepackage{helvet}
\usepackage{units}
\usepackage{multirow}
\usepackage{cite}

\usepackage{mathrsfs}
\usepackage{rotating}
\usepackage{xspace}
\usepackage{url}
\usepackage{slashed}

\newcommand{\RP}{\ensuremath{R}-parity\xspace}
\newcommand{\lam}[1][ijk]{\ensuremath{\lambda_{#1}}}
\newcommand{\lamp}[1][ijk]{\ensuremath{\lambda^{'}_{#1}}}
\newcommand{\lamdp}[1][ijk]{\ensuremath{\lambda^{''}_{#1}}}
\newcommand{\bilin}[1][i]{\ensuremath{\epsilon_{#1}}}

\newcommand{\LLE}{\ensuremath{LLE}\xspace}
\newcommand{\LQD}{\ensuremath{LQD}\xspace}
\newcommand{\UDD}{\ensuremath{UDD}\xspace}
\newcommand{\bRPV}{\ensuremath{b\text{RPV}}\xspace}

\newcommand{\meff}{\ensuremath{m_{\mathrm{eff}}}}
\newcommand{\mt}{\ensuremath{m_\mathrm{T}}}

\newcommand{\MJ}{\ensuremath{{M_J^\Sigma}}}

\newcommand{\ST}{\ensuremath{S_\mathrm{T}}\xspace}
\newcommand{\tauh}{\ensuremath{\tau_\mathrm{h}}\xspace}

\newcommand{\BR}[1]{\ensuremath{\mathrm{BR}(#1)}}

\newcommand{\BRb}[1][]{\ensuremath{\BR{b}}}
	
\newcommand{\mm}{\ifmmode{\mathrm{\ mm}}\else
                          \textrm{mm}\fi}
\newcommand{\mum}{\ifmmode{\mathrm{\ }\mu\mathrm{m}}\else
                           $\mu$\textrm{m}\fi}

\newcommand{\pythia}{{\sc Pythia}\xspace}

\newcommand{\herwigpp}{{\sc HERWIG++}\xspace}

\newcommand{\figref}[1]{Fig.~\ref{#1}}

\newcommand{\tabref}[1]{Tab.~\ref{#1}}

\newcommand{\Eqref}[1]{Eq.~\eqref{#1}} 

\newcommand{\beq}{\begin{equation}}
\newcommand{\eeq}{\end{equation}}
\newcommand{\bal}{\begin{align}}
\newcommand{\eal}{\end{align}}

\long\def\symbolfootnote[#1]#2{\begingroup\def\thefootnote{\fnsymbol{footnote}}
\footnote[#1]{#2}\endgroup}

\title{Searches for Prompt $R$-Parity-Violating Supersymmetry at the
LHC\footnote{Published 
in Advances in High Energy Physics, vol. 2015, Article ID 982167, 24 pages, 2015.}
}

\usepackage{authblk}

\author{Andreas Redelbach}

\affil{Julius-Maximilians-Universit\"at, W\"urzburg}

\abstracttext{ 
Searches for supersymmetry (SUSY) at the LHC frequently assume the conservation of \RP in their design, 
optimization and interpretation.
In the case that \RP is not conserved, constraints on SUSY particle masses tend to be weakened
 with respect to \RP-conserving models.
We review the current status of searches for \RP-violating (RPV)
  supersymmetry models at the ATLAS and CMS experiments, limited to 8~\tev\ search results published or submitted for 
publication as of the end of March 2015.
All forms of renormalisable RPV terms leading to prompt signatures have been considered in the set of analyses under review.
Discussing results for searches for prompt \RP-violating SUSY signatures summarizes the main constraints
 for various RPV models from LHC Run I and also defines the basis for promising signal regions to be optimized for
 Run II. 
In addition to identifying highly constrained regions from existing searches, also gaps in the
coverage of the parameter space of RPV SUSY are outlined.
}

\begin{document}

\tableofcontents
\clearpage

\section{Introduction}
\label{sec:Introduction}

One of the primary objectives of the detectors at the LHC is the  search for new particles and phenomena not described by the Standard Model of particle physics (SM).
Weak-scale supersymmetry~\cite{Miyazawa:1966,Ramond:1971gb,Golfand:1971iw,Neveu:1971rx,Neveu:1971iv,Gervais:1971ji,Volkov:1973ix,Wess:1973kz,Wess:1974tw} is a well-motivated and well-studied example of a theory beyond the SM (BSM) used to guide many of these searches.
One attractive feature of SUSY
is that it can solve the SM hierarchy problem~\cite{Dimopoulos:1981zb,Witten:1981nf,Dine:1981za,Dimopoulos:1981au,Sakai:1981gr,Kaul:1981hi}
 if the gluino, higgsino and top squark
masses are not much higher than the TeV scale.
Closely related to this is the paradigm of naturalness, see, for example,~\cite{Barbieri:1987fn,deCarlos:1993yy}.

In this document, we review constraints on SUSY models in the presence of lepton- or baryon-number violating interactions ($\slashed{L}$ and $\slashed{B}$, respectively) at the end of LHC Run I.
These interactions are present in generic SUSY models with minimal particle content.
They are renormalizable and are described by the following superpotential terms:
\begin{subequations}
\label{eq:rpvpotential}
\begin{equation}
\label{eq:LRPV}
W_{\slashed{L}\text{RPV}} =
\frac{1}{2}\lam L_iL_j\bar{E_k} +
\lamp L_iQ_j\bar{D_k} +
\bilin L_iH_2,
\end{equation}
\begin{equation}
\label{eq:BRPV}
W_{\slashed{B}\text{RPV}} =
\frac{1}{2}\lamdp \bar{U_i}\bar{D_j}\bar{D_k}.
\end{equation}
\end{subequations}
In this notation, $L_i$ and $Q_i$ indicate the lepton and quark SU(2)-doublet superfields, respectively, while $\bar{E_i}$, $\bar{U_i}$ and $\bar{D_i}$ are the corresponding singlet superfields. The indices $i$, $j$ and $k$ refer to quark and lepton generations.
The Higgs SU(2)-doublet superfield $H_2$ contains the Higgs field that couples to up-type quarks.
The \lam, \lamp and \lamdp{} parameters are new Yukawa couplings, referred to as
\emph{trilinear} $R$-parity-violating couplings.  
The \bilin{} parameters have dimensions of mass and are present in models with \emph{bilinear} $R$-parity violation
(\bRPV).
The terms in \Eqref{eq:rpvpotential} are forbidden in many models of SUSY by the imposition of \RP conservation 
(RPC)~\cite{Fayet:1976et,Fayet:1977yc,Farrar:1978xj,Fayet:1979sa,Dimopoulos:1981zb} in order to prevent rapid proton decay.
However, proton decay can also be prevented by suppressing only one of $W_{\slashed{L}\text{RPV}}$ or $W_{\slashed{B}\text{RPV}}$, in which case some \RP-violating interactions remain in the theory.

Introducing RPV couplings in the minimal supersymmetric Standard Model (MSSM) can signi\-ficantly weaken mass and
cross-section limits from collider experiments and also provide a rich phenomenology, see, for example, 
the articles~\cite{Barbier:2004ez,Allanach:2012vj,Asano:2012gj} or~\cite{1991NuPhB.365597D,Dreiner:2012wm}. 
A systematic phenomenological overview of possible signatures for specific RPV scenarios is summarized in Ref.~\cite{Dreiner:2012wm} going through all possible mass orderings and determining the dominant decay signatures.
Many papers have investigated signatures beyond the focus of most searches for SUSY at the LHC. Among such challenging scenarios are highly collimated LSP decay products~\cite{Graham:2014vya, Brust:2014gia}, same-sign dilepton signatures~\cite{Berger:2013sir,Saelim:2013gea},  
taus and $b$-jets with reduced missing transverse energy~\cite{Evans:2013uwa},
resonances of di-jets~\cite{Brust:2012uf, Franceschini:2012za},
high object multiplicities~\cite{Evans:2013jna}
or, more specifically, a charged lepton plus multiple jets~\cite{Lisanti:2011tm}.

In this note, we review the current constraints from various ATLAS and CMS searches for SUSY based on approximately 20~\ifb of $pp$ collision data with $\rts=8\tev$ collected in 2012.
This review is organized as follows:
After a short overview of \RP-violating parameters and previous constraints of RPV SUSY in
Section~\ref{sec:preATLAS}, the main characteristics of analyses searching for RPV SUSY at ATLAS and CMS
are presented in Section~\ref{sec:analyses_overview}.
The next sections focus on the results assuming the dominance of particular \RP-violating couplings:
After the results for \bRPV scenarios in Section~\ref{sec:bRPV}, several limits for simplified models
assuming \LLE, \LQD or a combination of \LQD and \LLE relevant for resonance production are discussed in
Sections~\ref{sec:LLE},~\ref{sec:LQD} and~\ref{sec:Resonance}, respectively.
In order to constrain models based on \UDD couplings, Section~\ref{sec:UDD} summarizes several results both from ATLAS and CMS searches.   
Finally, conclusions from \RP-violating searches at LHC Run I are drawn and some implications for strategies to investigate uncovered parts of the RPV SUSY parameter space for Run II are outlined.

\section{\RP-violating parameters and constraints}
\label{sec:preATLAS}
For each particle, $R$-parity is defined as $P_R = (-1)^{3(B-L)+2s}$ in terms of the corresponding spin, baryon and leptons numbers.
All Standard Model particles and the Higgs bosons have even $R$-parity, while all supersymmetric particles (sparticles)
have odd $R$-parity.   
As described, for example, in~\cite{Martin:1997ns}, an extension of the minimal supersymmetric Standard Model with \RP-violating interactions, does not extend the number of the supersymmetric particles.
Direct phenomenological consequences of \RP-violating interactions are:
\begin{itemize}
\item The lightest supersymmetric particle (LSP) is not necessarily stable.
\item Sparticles can also be produced in odd numbers, in particular single-sparticle production is possible.
\end{itemize}
Conversely, in \RP-conserving models, only pair production of SUSY particles is possible in collision processes, with the stable LSP being a possible candidate for dark matter.
An excellent review of LHC Run I searches with one focus on RPC SUSY is given by~\cite{Halkiadakis:2014qda}. 
In this section a short overview of RPV parameters and also of constraints previous to LHC searches are given.

\subsection{Parameters for RPV SUSY}
\label{sec:RP_parameters}
The number of  \RP-violating parameters can be obtained from~\Eqref{eq:LRPV} and~\Eqref{eq:BRPV}: Counting the possible generation indices in the terms $\bilin$ and $\lamp$ leads to 3 and 27 parameters, respectively.
As explained, for example, in~\cite{Barbier:2004ez}, antisymmetries in the summation over gauge indices,
suppressed in the notation of~\Eqref{eq:LRPV} and~\Eqref{eq:BRPV}, lead to  $\lam[ijk]=-\lam[jik]$ and  $\lamdp[ijk]=-\lamdp[ikj]$.
Due to these antisymmetric relations, 9 independent \RP-violating parameters of type \LLE and \UDD arise, respectively. 
The structure of trilinear \RP-violating couplings leads to Feynman diagrams as illustrated in~\figref{fig:diag_trilinearRPV} from~\cite{Barbier:2004ez}.
\begin{figure}
\centering
\includegraphics[width=0.8\textwidth]{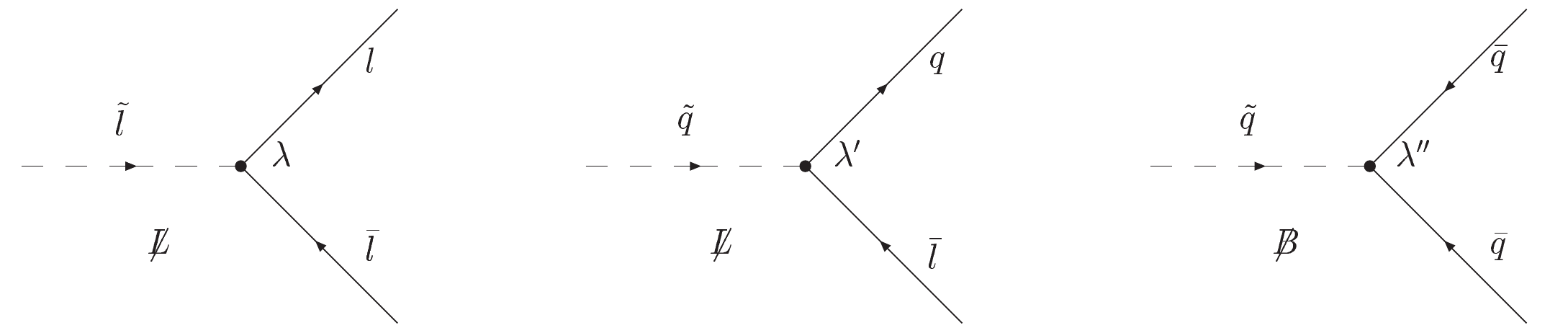}
   \caption{Feynman diagrams associated with the trilinear \RP-violating superpotential interactions
involving $\lam[]$, $\lamp[]$, or $\lamdp[]$.  $(\tilde{q}) q$ and $(\tilde{l}) l$ denote (s)quarks and
(s)leptons, respectively. Arrows on the (s)quark and (s)lepton lines are displayed to indicate the flow of the baryon or
lepton number.
      }
   \label{fig:diag_trilinearRPV}
\end{figure}

Possible signatures implied in various RPV scenarios are summarized in~\cite{Dreiner:2012wm}.
Although the majority of \RP-violating analyses focuses on neutralino (\ninoone) LSPs, alternative types of LSPs have been studied in~\cite{Dreiner:2009fi} and within the framework of \bRPV models also in~\cite{Hirsch:2003fe}.
It is also interesting to note that constraints for RPV couplings from theoretical considerations have been discussed, for example, in~\cite{Barbier:2004ez}:
In contrast to fixing individual RPV couplings explicitly, the assumption of spontaneous breaking of $R$-parity can lead to high predictivity for the actual values of these couplings.
One possible mechanism is based on right-handed sneutrino fields aquiring vacuum expectation values thus generating RPV couplings, see~\cite{Barbier:2004ez} and references therein.
In this context also the $\mu\nu$SSM~\cite{LopezFogliani:2005yw} as a natural extension of~\Eqref{eq:rpvpotential} is relevant, leading to interesting implications for LHC signatures, as recently discussed in~\cite{Mitsou:2015eka}.
Constraints for RPV couplings can also be derived from flavor symmetries, investigating, for example, flavor symmetry groups related to the Yukawa couplings and hierarchy of fermion masses. 
Within the so-called minimal flavor violation model~\cite{Csaki:2011ge}, the size of the small \RP-violating terms is determined by flavor parameters, and in the absence of neutrino masses only the \UDD terms remain in the superpotential~\Eqref{eq:BRPV}.
Recently, implications of fundamental symmetries on \RP-violation have been reviewed  in~\cite{Mohapatra:2015fua}, emphasizing, for example, that the simplest supersymmetric theories based on local $B-L$ predict that $R$-parity must be a broken symmetry.

\subsection{Pre-LHC constraints}
\label{sec:PreLHC_constraints}

A very large number of bounds for the trilinear \RP-violating couplings have been
deduced from studies of low and intermediate energy processes.
In particular, rare decays involving flavor violation, constitute
 constraints on RPV couplings.
Processes that violate lepton number or baryon number also
provide strong limits on \RP-violating couplings.
Presenting these indirect bounds is beyond the scope of this review, referring the reader to 
corresponding reviews, see, for example,~\cite{Barbier:2004ez} and~\cite{Kao:2009fg}. 
Many indirect bounds on the trilinear couplings assume \emph{single  
coupling dominance}, where a single \RP-violating coupling dominates over all the others.

However, it is important to note in general these bounds on \RP-violating couplings
are relaxed for increased masses of SUSY particles involved; see, for example,~\cite{Dreiner:1997uz}. 
In this context it is illustrative to look at the strong constraint derived from non-observation of
proton decay; see~\cite{Hinchliffe:1992ad}:
\begin{eqnarray}
  \lamp[11k]\cdot\lamdp[11k]\; \lesssim\; 10^{-23} \left( \frac{m_{\tilde{q}}}{100~\gev} \right)^2,
  \label{eq:rpvandudd:protondecay}
\end{eqnarray}
\noindent where $m_{\tilde{q}}$ is the typical squark mass.
As already noted before, it is sufficient to eliminate only one of $W_{\slashed{L}\text{RPV}}$ or
$W_{\slashed{B}\text{RPV}}$, corresponding to~\Eqref{eq:LRPV} or~\Eqref{eq:BRPV}, to forbid proton decay.
The form of the above constraint also shows the anti-correlation between the mass scale of intermediate SUSY particles and the size of their RPV couplings.
It is interesting to note that a possible solution to circumvent proton decay has also been established in the minimal flavor violation model~\cite{Csaki:2011ge}. 

Various results of analyses searching for non-vanishing \RP-violating couplings have been obtained in pre-LHC collider experiments. Constraints from existing searches for RPV SUSY can be classified, for example, by the category of contributing RPV couplings:
\begin{itemize}
\item LEP results~\cite{Abdallah:2003xc, Abbiendi:2003rn, Heister:2002jc, Achard:2001ek, Abdallah:2003xe} have investigated various \emph{trilinear RPV} couplings, typically leading to mass limits at the scale of 100~\gev.
\item HERA searches have mainly focused on signatures from \lamp couplings assuming the dominance of a single coupling~\cite{Aid:1996iw, Adloff:2001at, Boscherini:2002vb, Schneider:2002uf}. 
A very distinctive signature from a narrow-width resonance in  sparticle production with subsequent decay can be possible for a non-zero \LQD coupling or a combination of \LQD and \LLE couplings.
Therefore couplings of \lamp-type at HERA would allow resonant single squark production, corresponding to high sensitivities in different search channels. 
Expressing the limits in terms of the sparticle masses, squark masses above 200~\gev\ have been excluded.   
\item Several searches at the Tevatron~\cite{Allanach:1999bf}
have constrained various \emph{trilinear couplings} and/or sparticle masses even stronger.
The signatures studied at the Tevatron include searches for
multi-leptons~\cite{Abazov:2006nw,Abulencia:2007mp} (via \LLE) or multi-jets~\cite{Aaltonen:2011sg} and
pairs of di-jets~\cite{Aaltonen:2013hya} (via \UDD, respectively). 
Also resonant sparticle production~\cite{Tu:2008va,Aaltonen:2010fv,Abazov:2010km} 
 with subsequent decay (via a combination of \LQD and \LLE couplings) has been considered.
\end{itemize}
Since several of these signatures investigated at the Tevatron have set the strongest collider-based RPV limits before the LHC, we shortly mention some of these limits as benchmarks in comparison to LHC constraints to be discussed later.
Using the multi-jet final state, the CDF Collaboration has excluded gluino masses up to approximately 150~\gev\ for light-flavor models~\cite{Aaltonen:2011sg}. 
Based on the search for pairs of di-jets, as predicted from decays in stop-quark pair production, stop masses up to 100~\gev\ have been excluded~\cite{Aaltonen:2013hya}.
The CDF experiment has also set a limit on the expected cross-section at approximately 100~fb from multi-lepton search results~\cite{Abulencia:2007mp}.
In a benchmark scenario of resonant sneutrino production and subsequent lepton-flavor violating (LFV) decay
into different charged lepton flavors, $\tau$-sneutrino masses around 500~\gev\ have been excluded~\cite{Aaltonen:2010fv}.

Prior to LHC searches, no direct exclusion limits from LEP, HERA or Tevatron have been obtained for \bRPV models. 
However, several studies have investigated \bRPV phenomenology at the 
Tevatron~\cite{Allanach:1999bf,Datta:1999xq,Magro:2003zb,deCampos:2005ri}, elaborating, for example, signatures of multi-leptons or displaced vertices.

Significantly reducing the size of \RP-violating couplings generically leads to late LSP decays.  
Since the corresponding part of the RPV parameter space for small \RP-violating couplings does not predict prompt signatures at the LHC, no further details used in the searches for late decays are discussed here. 
It should however be mentioned that a number of analyses at ATLAS~\cite{ATLAS:2014fka,ATLAS-CONF-2013-092,Aad:2014gfa,Aad:2013yna,Aad:2013gva} 
and CMS~\cite{Chatrchyan:2013oca,Chatrchyan:2012jwg,Khachatryan:2014mea,CMS:2014wda,Khachatryan:2015jha} 
have probed signatures related to long-lived sparticles and displaced vertices expected in such cases deriving also strong limits on SUSY masses. 
A phenomenological overview of these searches for long-lived sparticles has recently been presented, for example, in~\cite{Liu:2015bma}. 

\section{Overview of analyses searching for RPV SUSY}
\label{sec:analyses_overview}

Both ATLAS~\cite{Aad:2008zzm}  and CMS~\cite{Chatrchyan:2008aa} are multi-purpose detectors 
designed for the study of $pp$ and heavy-ion collisions at the LHC. 
They provide nearly
full solid angle coverage around the interaction point.
Each detector uses a
  right-handed coordinate system with its origin at the nominal
  interaction point in the centre of the corresponding detector 
and the $z$-axis  along the beam pipe. 
Cylindrical coordinates
  ($r$,$\phi$) are used in the transverse plane, $\phi$ being the azimuthal angle
  around the beam pipe. 
The pseudorapidity is defined in terms of the
  polar angle $\theta$ as $\eta = - \ln \tan(\theta/2)$.

\subsection{Strategies for simulating and selecting events}
\label{sec:sim_select}
The 8~\tev~$pp$ data set, after the application of beam, detector and
data quality requirements, has an integrated luminosity of approximately $20$
\ifb~both for ATLAS and CMS detectors. 
It is interesting to note that the average number of $pp$ interactions occurring in the
same bunch crossing at 8~TeV varies between approximately 10 and 30, necessitating systematic studies of related pile-up effects. 
The trigger system of ATLAS and CMS 
consists of hardware-based systems, with subsequent software-based systems. 
Using so-called High Level Triggers, the events of interest are finally recorded.
For each analysis, a combination of different triggers is used, before the offline selection of events is done. 
The main requirements of the latter are summarized for each analysis in Sections~\ref{sec:details_searches_A} and~\ref{sec:details_searches_C}.
In order to ensure the quality of reconstruction,
various requirements on transverse momenta \pt\ and criteria for the isolation to other objects have been developed at the LHC detectors, 
with more details presented in ~\cite{Aad:2008zzm}  and~\cite{Chatrchyan:2008aa}. 
After reconstruction of final states, the most relevant physical objects for prompt RPV SUSY analyses can shortly be classified as follows:

Electrons, muons, hadronically decaying $\tau$-leptons, collectively referred to as charged leptons.
Depending on the specific analysis, it is possible to discriminate hadronic jets according to their flavor contents:
In particular, $b$-tagged jets can often be distinguished from jets consisting of only light quark-flavors.
The missing transverse energy per event  \met\ is computed using the transverse momenta of identified objects.

SUSY \RP-violating signal samples are generated using different event generators, for example,
\herwigpp~\cite{Bahr:2008pv} or \pythia~\cite{Sjostrand:2006za,Sjostrand:2007gs}.
The events are subsequently simulated within the framework of fast or full simulation, where, 
for the details of the specific setup of event generation or simulation, the corresponding analysis papers should be considered.

Unless otherwise stated, signal cross-sections are calculated to next-to-leading order in the strong coupling constant, adding the resummation of soft gluon emission at next-to-leading-logarithmic accuracy (NLO+NLL)~\cite{Beenakker:1996ch,Kulesza:2008jb,Kulesza:2009kq,Beenakker:2009ha,Beenakker:2011fu}.
The nominal cross-section and the uncertainty are taken from an envelope of cross-section predictions using different PDF sets and factorisation and renormalisation scales, as described in Ref.~\cite{Kramer:2012bx}.

Each analysis is based on a number of \emph{signal regions} (SRs), each designed to maximize the sensitivity to different final state topologies
in terms of the chosen discriminating variables. 
Additionally, a number of control regions (CRs) are constructed to
constrain the dominant backgrounds. 
These control regions are designed to
have a high purity and a small statistical uncertainty in terms of the background process
of interest and also to contain only a small fraction of the potential SUSY signal. 
Practically, control regions are often introduced to estimate the rate of SM processes, using data-driven methods or
also normalization of Monte Carlo simulations. 

\subsection{Strategies for presentation of results}
\label{sec:presentation_results}

The large number of free mass parameters for sparticles in the MSSM is already severely constrained by many experimental bounds, see, for example, the discussion in~\cite{Martin:1997ns}. 
As a consequence, several approaches to study SUSY particle spectra have been developed:

Within the phenomenological MSSM \emph{pMSSM}~\cite{AbdusSalam:2009qd}, the high number of free SUSY parameters is reduced with realistic requirements on the flavor and
CP structure, without imposing any SUSY-breaking scheme.
In this framework also SUSY spectra consistent with various experimental constraints, as, for example, the LHC results for the Higgs mass can be addressed~\cite{Arbey:2012bp,Dumont:2013npa}.  
 
The approach of \emph{simplified models}~\cite{Alwall:2008ag,Alves:2011wf} is commonly used in searches for SUSY at the LHC. In this case
the decay cascades are modeled simply by setting the masses of most SUSY particles to multi-TeV values, 
effectively decoupling them for the reach at the LHC.
This also implies selection of specific production channels, while other mixed production modes, for example,
scalar plus fermionic SUSY particle are typically neglected. 
The decay cascades of the remaining particles to the LSP, typically with zero
or one intermediate step, are characterized only by the masses of the participating particles,
allowing studies of the search sensitivity to the SUSY masses and decay kinematics. 

In an alternative approach, \emph{complete SUSY models} as, for example,  mSUGRA/CMSSM~\cite{Chamseddine:1982jx,Barbieri:1982eh,Ibanez:1982ee,Hall:1983iz,Ohta:1982wn,Kane:1993td} or 
minimal GMSB~ \cite{Dine:1981gu,AlvarezGaume:1981wy,Nappi:1982hm,Dine:1993yw, Dine:1994vc,Dine:1995ag},
 are simulated. 
These models typically
impose boundary conditions at a high energy scale and determine the SUSY masses near the \tev\ scale 
by evaluating renormalization group equations. 
Due to the minimal number of input parameters at the high energy scale,  it is realistic to scan the parameter space effectively. 

One common strategy for obtaining results is to compute
the level of agreement between the background prediction and data using 
the $p$-value for the number of observed events to be consistent with the
background-only hypothesis.
To do so, the number of events in each signal region is described using a Poisson
probability density function. The statistical and systematic uncertainties
on the expected background values are modeled with nuisance parameters
constrained by a Gaussian function with a width corresponding to the size of the uncertainty considered. 

Since no significant excess of events over the SM expectations is observed
in any signal region of the \RP-violating analyses,
upper limits at 95\% CL on the number of BSM events for each signal region can be derived in a \emph{model-independent} way.
Here the CL$_s$ prescription~\cite{Read:2002hq} is used. 
Normalising these events by the integrated luminosity $\mathcal{L}$ of the data sample, they can be interpreted
as upper limits on the visible BSM cross-section ($\sigma_{\rm vis}$), where
$\sigma_{\rm vis}$ is defined as the product of acceptance, reconstruction
efficiency and production cross-section. 
If a limit on non-SM events ($N_\text{non-SM}$) has been obtained in a BSM analysis, the visible signal cross-section can also be determined as $\sigma_\text{vis} = N_\text{non-SM}/{\cal{L}}$.

\emph{Model-dependent limits} will be discussed in detail in sections~\ref{sec:bRPV} to~\ref{sec:UDD}.
For many models, the limits are calculated from asymptotic formulae~\cite{Cowan:2010js}
with a simultaneous fit to all signal regions based on the profile likelihood method.
Alternatively, the limit can also be obtained from pseudo experiments; further details can be found in each paper.

The systematic uncertainties on the signal expectations
originating from detector effects and the theoretical uncertainties on the signal acceptance are included. 
The impact of the theoretical uncertainties on the signal cross-section is shown on the limit plots obtained.
The $\pm1 \sigma^{\rm SUSY}_{\rm theory}$ lines around the observed limits are
obtained by changing the SUSY cross-section by one standard deviation ($\pm1\sigma$).
All mass limits on supersymmetric particles quoted later are derived from 
the $-1 \sigma^{\rm SUSY}_{\rm theory}$ theory line.
The band around the expected limit shows the $\pm1\sigma$ uncertainty,
including all statistical and systematic uncertainties except the theoretical uncertainties
on the SUSY cross-section. 
If several SRs contribute to exclusion limits in a model investigated, the general strategy is to obtain
 limits by performing a statistical combination
of the most sensitive signal regions.

\subsection{Details for RPV searches at ATLAS}
\label{sec:details_searches_A}
In this section the main requirements for signal selections developed in ATLAS searches for \RP-violating SUSY are summarized, also introducing relevant kinematic variables.
Some of these analyses have been optimized also for RPC scenarios, however the focus of this review is on RPV-related signal regions. 
Each of these analyses has investigated RPV-related constraints in at least two different RPV models 
or several SRs have been developed particularly for RPV signatures.\footnote{When finalizing this
review, a recent ATLAS analysis~\cite{ATLAS-CONF-2015-015} has also considered stop \LQD-type decays to
light charged lepton plus
$b$-quark, constraining stop masses up to 1~TeV.}  
\begin{itemize}
\item {\bf Multi-lepton analysis (ATLAS)}
In this case at  least four charged leptons in every signal event are required, at least two of which must be electrons or muons, in the following referred to as ``light leptons''.
The events are separated into signal regions based on the number of light leptons observed~\cite{Aad:2014iza}, and
the absence of a \Zboson\ boson candidate among the pairs of light leptons facilitates suppression of backgrounds in \RP-violating searches.
The SM background is further reduced using the missing transverse momentum \met\ and the effective mass \meff,
defined in this case as the scalar sum of the \met, the \pt\ of all selected charged leptons
and the \pt\ of reconstructed jets. 
In most signal regions, events with a pair of light leptons forming a \Zboson\ boson candidate are vetoed,
and possible $\Zboson\to\ell^+\ell^-\gamma$ and $\Zboson\to\ell^+\ell^-\ell^+\ell^-$ candidates
are also rejected.
Three signal regions are based on threshold requirements only for \met, thus being useful in particular for RPC searches for low sparticle masses.  
Additionally, in the SRs used for RPV results, either high \met\ \emph{or} high \meff\ is required -- thus,
a selected event may have one quantity below the threshold, but never both.
As the SRs used have disjoint selection criteria,
they are statistically combined when setting constraints on the specific SUSY models considered in~\cite{Aad:2014iza}.

\item {\bf Same-sign/three-lepton analysis  (ATLAS)} 
The search~\cite{Aad:2014pda} requires two light leptons with same charge or three
light leptons in conjunction with requirements on the number of jets.
It designed in particular for SUSY models where pair-produced Majorana particles (for example, gluinos) can decay semileptonically with a large branching ratio.
The effective mass, \meff, is a key discriminating variable, defined by this analysis as the sum of \met\ and the \pt\
values of the signal leptons and all signal jets. 
If the event contains a third light lepton
the event is regarded as
three-lepton event, otherwise it is a two-lepton event.
Five non-overlapping signal regions have been defined in total. 
The signal regions SR3b and SR1b use leptons,  large \meff\  and also the presence of $b$-jets to suppress the SM background.
There is no explicit \met\ requirement in SR3b, implying that this SR does not depend on the assumption of a stable LSP escaping the detector unseen.
SR1b additionally uses the transverse mass, \mt, to reject background events with \Wboson\ bosons, defined as
\begin{equation}
\mt =
\sqrt{2 \pt^{l} \met
 (1-\cos[\Delta\phi(\mathbf{l},\mathbf{p}_{\mathrm{T}}^\mathrm{miss})])},
\label{eq:mT}
\end{equation}
where $\pt^{l}$ is the larger of the \pt\ values of the two charged leptons, and $\mathbf{p}_{\mathrm{T}}^\mathrm{miss}$ is the missing transverse momentum vector.

\item {\bf Tau plus jets analysis (ATLAS)}
Requiring at least one tau lepton in events with jets and large \met\, the
search~\cite{Aad:2014mra} can also be sensitive to RPV models with relatively
high multiplicities of taus.
The search channels are separated by the numbers of taus and light charged leptons involved, leading to
$e\tau$, $\mu\tau$ and $\tau\tau$ channels, respectively.
The following kinematical variables are introduced to suppress background processes:
The transverse mass formed by \met\ and the \pt\  of the tau lepton in the $e\tau$ and $\mu\tau$ channels,
\begin{equation}
m^\tau_{\textrm{T}}=\sqrt{2\pt^{\tau}\met(1-\cos(\Delta\phi(\mathbf{\tau},\mathbf{p}_{\mathrm{T}}^\mathrm{miss})))}.
\end{equation}
Similarly the transverse mass $m^l_{\textrm{T}}$ formed by \met\ and the \pt\ of the light lepton ($e$ or $\mu$) is used.
Two variants of $\HT$-related variables have been defined as the scalar sum of the transverse momenta of the tau, light lepton and signal jets: $\HT$ includes all signal jet ($\pt>$30~GeV) candidates, whereas  $\HT^{2j}$ only considers two jets with the largest transverse momenta in the event.
In this analysis the effective mass uses $\HT^{2j}$, that is, $\meff=\HT^{2j}+\met$.
Moreover a requirement on the minimal azimuthal angle $\Delta\phi(\text{jet}_{1,2},\pt^{\textrm{miss}})$
between $\pt^{\textrm{miss}}$ and either of the two leading jets is used to remove multi-jet events.
As a result, also upper limits on the visible cross-section have been obtained for the \bRPV-related SRs of type $e\tau$, $\mu\tau$ and $\tau\tau$, respectively. 

\item {\bf Multi-jet analysis (ATLAS)}
 Two complementary search strategies have been developed in this analysis~\cite{Aad:2015lea}:
 The \emph{jet-counting analysis} is searching for an excess in events with $\geq$6-jet or $\geq$7-jets, using the
 predictable scaling of the number of $n$-jet events ($n=6,7$) as a function of the transverse momentum (\pt)
 requirement placed on the $n$th leading jet in \pt\ for background processes.
It is interesting to note that this scaling relation differs significantly between the signal and the background. 
 That analysis technique provides the opportunity to enhance sensitivity to specific heavy-flavor compositions in the final state and to explore various assumptions on the branching ratios of the benchmark signal processes studied.
The number of jets, the \pt\ requirement used in the selection of jets, and the number
of $b$-tagged jets are optimized separately for each signal model.

The second approach in~\cite{Aad:2015lea} consists of a data-driven template-based analysis using a topological observable called the \emph{total-jet-mass} of large-radius ($R$) jets. 
This analysis method is based on templates of the event-level observable formed by the scalar sum of the four leading large $R$ jet masses in the event, which is significantly larger for the signal than for the SM backgrounds.
The total-jet-mass analysis uses a topological observable $\MJ$ as the primary distinguishing characteristic between signal and background. 
The observable $\MJ$~\cite{Hook:2012fd, Hedri:2013pvl, Cohen:2014epa} is defined as the scalar sum of the masses of the four leading large-radius jets reconstructed with a radius parameter $R=1.0$, $\pT>100$~GeV and $|\eta|<\eta_{cut}$,

\begin{equation}
\MJ = \sum^4_{{\pT}>100\,\textrm{GeV} \atop |\eta|\leq \eta_{cut}} m_{jet} .
\label{eq:defMJ}
\end{equation}

As explained, for example, in~\cite{Aad:2015lea}, four-jet (or more) events are used, because four large-$R$ jets cover a significant portion of the central region of the calorimeter, and are very likely to capture most signal quarks within their area. 
As a second discriminating variable for the design of SRs and CRs the pseudorapidity difference $|\Delta \eta|$ between the two leading large-$R$ jets is used. 
This is motivated by different angular distributions among jets expected from signal events as compared
to background processes.
For the definition of SRs also the \pt\ thresholds of the third and fourth jet have been included. 
Using the results from simulation studies, it has been demonstrated that $\MJ$ typically has higher sensitivity than the kinematic variable \HT. 
The latter is essentially a measure of the transverse energy (or transverse momenta) in the event, whereas the $\MJ$ mass intrinsically also contains angular information to be used in high-multiplicity jet events.   
This analysis technique focuses primarily on the ten-quark models as further discussed in section~\ref{sec:UDDmulti-jet_models_ATLAS}.
The total-jet-mass analysis is designed to be independent of the flavor composition of the signal process and as a data-driven method it essentially removes any reliance on MC simulations of these hadronic final states. 
No explicit veto is applied to events with leptons or \met.

Also model-independent upper limits on non-SM contributions have been derived separately for each analysis in~\cite{Aad:2015lea}.

\item {\bf LFV resonance analysis (ATLAS)}
The reconstruction of a narrow-width resonance from its decay products essentially relies on the invariant mass determined from the corresponding momenta.
In this case the decay products are given by charged leptons of different flavor~\cite{Aad:2015pfa}.
Therefore the selection for signal events requires exactly two leptons ($l^+_i l^-_j$ with $i\neq j$), of opposite charge and of different flavor. 
Good discrimination against background is obtained requiring that
the two leptons are back-to-back in the azimuthal plane with 
$|\Delta \phi_{\ell\ell'}|>2.7$, where $\Delta \phi_{\ell\ell'}$ is the $\phi$ difference between 
the two leptons.
In events containing a hadronically decaying $\tau$, it is additionally required that the transverse energy $E_T$ of the $\tau$ candidate is less than the corresponding $E_T$ of the light signal lepton due to the energy carried by the $\tau$-neutrino. 

In order to reconstruct the four-momenta of hadronically decaying $\tau$-leptons, also the momenta of the emerging $\tau$-neutrino have to be taken into account.
A collinear neutrino approximation is used to determine the dilepton invariant mass ($m_{ij}$) in the $e\tau_{had}$ and $\mu\tau_{had}$ channels. 
This approximation is well justified since the hadronic decay of a high-energetic $\tau$ lepton from a heavy resonance, the neutrino and the resultant jet are 
nearly collinear. 
The four-vector of the neutrino is reconstructed from the $\vec{p}_{\text T}^{\hspace{1.5pt}\text {miss}}$
and $\eta$ of the $\tau_{had}$ jet.
Four-vectors of the electron or muon, $\tau_{had}$ candidate and neutrino are 
then used to calculate the \emph{dilepton invariant mass} $m_{ij}$. 
The minimal requirement on $m_{ij}$ for signal events is $m_{ij}>200$~\gev.
Finally, the expected and observed upper limits are obtained as a function of  $\tilde{\nu}_{\tau}$ mass. 

\end{itemize}

For further reference, the ATLAS analyses considered are summarized in Table~\ref{tab:analyses_overview_prompt_ATLAS} indicating the signatures and main variables for signal selections.
An overview relating ATLAS analysis and corresponding RPV model investigated, is presented in Table~\ref{tab:analyses_overview_model_ATLAS}. 
\begin{table}[tbp]
\centering
\begin{tabular}{c|ccc}
\hline
 Short name              & Signature                                                  &  Variables                                                                                 & Ref. \\ 
 \hline
 4L$_A$                     &   $4(e, \mu, \tau) + \met$                        &  $ \meff, \met$, $Z_{\textrm{veto}}$                                  & \cite{Aad:2014iza} \\
 SS/3L$_A$                &   $\ell^{\pm}\ell^{\pm}$ or $3\ell$            &  $N_{jets}$, $N_{b-jets}$, $\met$, $\meff $                                & \cite{Aad:2014pda} \\
 $\tau_A$                  & $\tau +\met $ $\geq 4j$                 &  $m^\tau_{\textrm{T}}$, $m^l_{\textrm{T}}$, $\HT$, $\HT^{2j}$, $\met$, $\meff$, $\pt^{\textrm{miss}}$    & \cite{Aad:2014mra}  \\
 multi-j$_A$               &   $\geq 6j$                                              &  jet $\pt$, $M_J^\Sigma$,  $N_{b-jets}$, $|\Delta \eta|$     &  \cite{Aad:2015lea}  \\
l$_i$l$_j$ reson$_A$ &  $e\mu$, $e\tau$, $\mu\tau$ resonance  &  $|\Delta \phi_{ij}|$, $m_{ij}$  ($i\neq j$)                      &  \cite{Aad:2015pfa} \\ \hline
\end{tabular}
 \caption{Overview of ATLAS analyses designed to probe prompt RPV models.
The signature descriptions are indicative only, further details can be found in the analysis documentation  in each case.
\label{tab:analyses_overview_prompt_ATLAS}}
\end{table}

\begin{table}[tbp]
\centering
\begin{tabular}{ccccc}
\hline
RPV type     & RPV couplings                                                    & Production                           &  LSP &  Analysis  \\
 \hline
\LLE          & \lam[12k], \lam[k33]~($k=1,2$)      &        \chinoonepm , slepton, sneutrino, gluino   & \ninoone         &    4L$_A$  \\
\UDD        & \lamdp[323]                               &      Gluino                                                 &     \stopone          &     SS/3L$_A$  \\
\bRPV       &  \bilin~($i=1,2,3$)                       &        Strong/electroweak (mSUGRA)        &  \ninoone            &    SS/3L$_A$, $\tau_A$  \\
\UDD        &   \lamdp[ijk]                               &               Gluino                                       &   \ninoone                &      multi-j$_A$  \\
\LQD+\LLE  &  $\lamp[311]$ and $\lam[i3j] $  ($i\neq j$)     &  $\tau$-sneutrino    &  $\tilde{\nu}_\tau$         &     l$_i$l$_j$ reson$_A$   \\
\hline 
\end{tabular}
 \caption{Schematic overview of RPV model parameters investigated by ATLAS analyses.
The signature descriptions are also indicated in \tabref{tab:analyses_overview_prompt_ATLAS} with corresponding references.
\label{tab:analyses_overview_model_ATLAS}}
\end{table}

\newpage
\subsection{Details for RPV searches at CMS}
\label{sec:details_searches_C}

An overview of CMS analysis strategies used in the search for prompt RPV is given below.

\begin{itemize}
\item {\bf Multi-lepton plus b-jets analysis (CMS)}
In this analysis, events with three or more charged leptons are selected, requiring two light leptons, which may be electrons or muons~\cite{Chatrchyan:2013xsw}. 
Accepting only opposite-sign, same-flavor pairs of electrons or muons with an invariant mass $m_{\ell \ell} > 12$~\gev\
reduces backgrounds from Drell-Yan processes and low mass resonances.
Signal regions are defined with different requirements on the total number of light leptons and the number of hadronically decaying $\tau$ candidates in the event. 
Since no \Zboson\ bosons are expected in the signal models under investigation, events in which any
selected dilepton pair has an invariant mass consistent with that of the \Zboson\ boson are rejected thus providing good suppression of \Zboson\-related backgrounds.
Moreover, at least one $b$-tagged jet is required in the signal regions.
Additional discrimination against background events is obtained with cuts on the $\ST$ distribution.
As was discussed in~\cite{Chatrchyan:2013xsw}, that distribution has high sensitivity to the mass of the parent particle, produced in pair production.

Several kinematic regions relevant for results are introduced in~\cite{Chatrchyan:2013xsw}, 
relating to different assumptions on stop masses in comparison to neutralino LSP masses. 
Relatively light stops (with respect to masses of $\ninoone$) would correspond to region A, while the case of heaviest stop masses (in comparison to $m(\ninoone)$) is included in region E. 
It is interesting to note that, for example, in region B stop four-body decays  $\stopone \rightarrow t \mu t \bar{b}$ or $ t \nu b \bar{b} $ are possible. 

\item {\bf SS-leptons analysis (CMS)}
The analysis~\cite{Chatrchyan:2013fea} targets at topologies with same-sign leptons and additional jets from strong production processes.
Events with at least two isolated same-sign leptons ($ee$, $e\mu$ or $\mu\mu$)
and at least two jets  are selected.
The lepton pairs are required to have an invariant mass above 8~GeV and also 
events with a third lepton are rejected if the lepton forms an
opposite-sign same-flavor pair with one of the first
two leptons for which the invariant mass of the pair ($m_{\ell \ell}$) satisfies
$m_{\ell \ell} < 12$~\gev\ or $76 < m_{\ell \ell} < 106$~\gev.
Signal regions are defined with different requirements on \met, \HT, the number of jets, and the number of $b$-tagged jets.

For each model considered, limits are obtained by performing a statistical
combination of the most sensitive signal regions.
 The search region dedicated to RPV results is based on the selection of  $n_{jets}\geq$ 2,
 $n_{b-jets}\geq$ 2, and 
 $\HT>500$~\gev\ without explicit requirement on \met~\cite{Chatrchyan:2013fea}.

\item {\bf  $\tau$ + $b$-jets analysis (CMS)}
Different assumptions for the decays of stops have motivated the search for signatures of $\tau$-leptons and $b$-jets~\cite{Khachatryan:2014ura}. 
Selected events are required to contain a light lepton and a hadronically decaying $\tauh$ of opposite electric charge thus leading to the signal channels
$e\tauh$ and $\mu\tauh$.
Events are vetoed if another light lepton is found, passing
the kinematic, identification, and isolation criteria,
which has an opposite electric charge from the selected light lepton.
The $b$-tagged jet with the highest \pt\ is selected, and then the remaining four jets
with the highest \pt\ are selected whether or not they are $b$-tagged.
The \ST distribution is finally used to extract the limits, where \ST is defined as the scalar
sum of the \pt\ of the light lepton, the $\tauh$, and the five jets.

\item {\bf  $\geq 4j$  from jet pairs analysis (CMS)}
This analysis~\cite{Khachatryan:2014lpa} has been designed to search for pairs of jets where each jet decays to two jets, respectively.
The strategy followed in this analysis first requires that signal events contain at least four jets.
The leading four jets, ordered in \pt, are used to create three unique combinations
of di-jet pairs per event. 
A distance variable is implemented to select
the jet pairing that best corresponds to the two resonance decays,
$\Delta R = \sqrt{(\Delta\eta)^2 + (\Delta\phi)^2}$, where
$\Delta\eta$ and $\Delta\phi$ are the differences in $\eta$ and $\phi$ 
between the two jets, respectively. 
This variable exploits the smaller
relative distance between daughter jets from the same parent
decays compared to that between uncorrelated jets. 
For each di-jet pair configuration the value of $\Delta R_{\text{di-jet}}$ is calculated:
\begin{equation}
\Delta R_{\text{di-jet}} = \sum_{i=1,2}|\Delta R^{i} - 1|,
\label{eqn:dRpairing}
\end{equation}
where $\Delta R^{i}$ represents the separation between two jets in di-jet pair $i$ and an
offset of 1 is used to maximize the signal efficiency. 
The configuration that minimizes the value $\Delta R_{\text{di-jet}}$ is selected,
with $\Delta R_{\text{min}}$ representing the minimum $\Delta R_{\text{di-jet}}$ for the event.

Once a di-jet pair configuration is chosen, two additional quantities
are used to suppress the backgrounds from SM multi-jet
events and incorrect signal pairings:
The pseudorapidity difference between the two di-jet systems $\Delta\eta_{\text{di-jet}}$, and the absolute value of the fractional mass difference $\Delta m/m_{\text{av}}$,
where $\Delta m$ is the difference between the two di-jet masses and $m_{\text{av}}$
is their average value. 
As discussed in~\cite{Khachatryan:2014lpa} the $\Delta m/m_{\text{av}}$ quantity is
small with a peak at zero in signal events where
the correct pairing is chosen, while for SM multi-jet background or
incorrectly paired signal events, this distribution is much broader. 
An additional kinematic variable $\Delta$ is calculated for each di-jet pair:
\begin{equation}
\Delta = \left(\sum_{i}^{1,2} |\pt^{i}|\right) - m_{\text{av}},
\label{eqn:delta}
\end{equation}
where the \pt\ sum is over the two jets in the di-jet configuration.
This type of variable has been used extensively in hadronic resonance
searches at the Tevatron and the LHC, see, for example,~\cite{Chatrchyan:2013izb} and references therein.
Requiring a minimum value of $\Delta$ results in a lowering of the
peak position value of the $m_{\text{av}}$ distribution from background SM g events.
With this selection the fit to the background can be extended to lower values of $m_{\text{av}}$, making
a wider range of supersymmetric particle masses accessible to the search~\cite{Khachatryan:2014lpa}.

\item {\bf Multi-jet analysis (CMS)}
This search targets jets final states with high multiplicities from pair-produced three-jet resonances~\cite{Chatrchyan:2013gia}.
Signal events have to contain at least six jets with additional requirements on \pt~thresholds. 
The {\it jet-ensemble technique}~\cite{Aaltonen:2011sg}
is used to combine the six highest-\pt~jets
in each event into all possible unique triplets.
To maximize sensitivity to
the presence of a three-jet resonance, an additional requirement
is placed on each jet triplet to suppress SM
backgrounds and remove incorrectly combined signal triplets. 
This selection criterion is based on the constant invariant mass of correctly reconstructed signal triplets and
also on the observed linear correlation between the invariant mass and scalar sum of jet \pt\ for background triplets and incorrectly combined signal triplets:
\begin{equation}
M_{jjj} < \left( \sum_{i=1}^{3}\pt^{i} \right)-\Delta \;,
\label{eq:diag_def}
\end{equation}
where $M_{jjj}$ is the triplet invariant mass,
\pt\ sum is over the three jets in the triplet (triplet scalar \pt), and $\Delta$
is an empirically determined parameter.
The peak position of the $M_{jjj}$ distribution in data depends on the value of $\Delta$,
where $\Delta=110~\gev$ is found to be the optimal choice, yielding the lowest value of the peak of $M_{jjj}$.

The use of $b$-jet identification facilitates a heavy-flavor search in addition to the
inclusive search for three-jet resonances.
High-mass signal events lead to a
more spherical shape than background events, which typically
contain back-to-back jets.
In order to significantly reduce the background in the
high-mass searches, a sphericity
variable, $S = \frac{3}{2} (\lambda_{2}+\lambda_{3})$ is used, where the
$\lambda_{i}$ are eigenvalues of the following tensor~\cite{Sjostrand:2006za}:
\begin{equation}
S^{\alpha \beta} = \frac{\sum\limits_{i} p^{\alpha}_{i}  p^{\beta}_{i}}{\sum\limits_{i} |\boldmath{p}_{i}|^2} \;
\label{eqn:sphericity_tensor}
\end{equation}
Here $\alpha$ and $\beta$ label separate jets, and the sphericity $S$ is calculated
using all jets in each event.
In summary, the SRs in this analysis are defined using $M_{jjj}$, $\Delta$ and also cuts on the fourth jet \pt, sixth
jet \pt\ and $\ST$.
To optimize sensitivity for the heavy flavor search, a region of low or high mass $M_{jjj}$ for the underlying resonance mass has been developed, respectively.  

\end{itemize}

This overview of CMS RPV analyses is completed with Tables~\ref{tab:analyses_overview_prompt_CMS} and~\ref{tab:analyses_overview_model_CMS}.
Using these tables, the signatures and main variables for signal selections per analysis are indicated and also the information which analyses are used to constrain which RPV SUSY models is presented.
As can be noted from Table~\ref{tab:analyses_overview_model_CMS}, most of these analyses from CMS have also investigated at least two different RPV-based models.  

\begin{table}[tbp]
\centering
\begin{tabular}{c|ccc}
\hline
 Short name              & Signature                                                   &  Variables                                                                             & Ref. \\ 
 \hline
 3L/b$_C$                  &   $\geq 3\ell + \textrm{ b-jets} $      &  $ S_T$, $m_{\ell \ell}$                            &    \cite{Chatrchyan:2013xsw}    \\  
 SS$_C$                     &      $\ell^{\pm}\ell^{\pm}$                &    $N_{jets}, N_{b-jets}, \HT$                 &  \cite{Chatrchyan:2013fea} \\ 
 $\tau$b$_C$            &   $\tau$ + \textrm{b- jets}               &  $N_e=1$ or $N_\mu=1$, $\ST$              &  \cite{Khachatryan:2014ura} \\      
pair-j$_C$                  &  $\geq 4j$  from jet pairs                &  $\Delta m$,
$\Delta\eta_{\text{di-jet}}$, $\Delta$, 4th jet \pt  &  \cite{Khachatryan:2014lpa}\\   
multi-j$_C$                &  $\geq 6j$                                         &  $\Delta$, 4th jet \pt, 6th jet \pt, $\ST$     &  \cite{Chatrchyan:2013gia} \\ 
\hline 
\end{tabular}
 \caption{Overview of CMS analyses constraining prompt RPV models.
The signature descriptions are indicative only, the reader is referred to the analysis documentation for further details in each case.
\label{tab:analyses_overview_prompt_CMS}}
\end{table}

\begin{table}[tbp]
\centering
\begin{tabular}{ccccc}
\hline
RPV type   & RPV couplings            & Production                           &  LSP &  Analysis  \\
 \hline
\LLE / \LQD  & \lam[122], \lam[233], \lamp[233] & Stop  &      \ninoone & 3L/b$_C$   \\
\UDD         & \lamdp[323]                                 &  Gluino     &  \stopone     & SS$_C$  \\
\LQD         &   \lamp[3jk]~($j,k=1,2$), \lamp[333]       & Stop   &  \ninoone, \stopone  & $\tau$b$_C$ \\
\UDD     & \lamdp[312],  \lamdp[323]  & Stop  &   \stopone  &  pair-j$_C$  \\
\UDD  & \lamdp[112],  \lamdp[113],  \lamdp[223] & Gluino  &    \ninoone   &multi-j$_C$  \\
\hline
\end{tabular}
 \caption{Overview of RPV model parameters investigated by CMS analyses.
The signature descriptions are also indicated in \tabref{tab:analyses_overview_prompt_CMS} with corresponding references.
\label{tab:analyses_overview_model_CMS}}
\end{table}

\section{Bilinear \RP violation}
\label{sec:bRPV}

In the \bRPV model, the terms with coefficients \bilin{} ($i=1,2,3$) lead to lepton-number violating
interactions between lepton and Higgs superfields. 
An overview of \bRPV phenomenology can be found, for example, in Refs.~\cite{Porod:2000hv} and~\cite{Diaz:2003as}.
Note that also for the soft SUSY breaking terms additional \bRPV terms $-B_i\bilin \tilde{L}_i H_2$ and
$m_{\ell_i H}^2\tilde{L}_i H_1^\dagger$~\cite{Allanach:2008qq} 
arise,
leading to extra parameters. 
In general, there is no basis where both sets of bilinear RPV terms
$\bilin L_iH_2$  and $B_i\bilin \tilde{L}_i H_2$ can be eliminated at the same time. 
Taking into account the mixing of sneutrinos and scalar neutral Higgs fields, the electroweak symmetry is broken
when these scalar fields acquire vacuum expectation values.
Another characteristic consequence of \bRPV is the generation of neutrino masses via
neutralino-neutrino mixing; see, for example,~\cite{Romao:1999up}.

Requiring both that electroweak symmetry breaking is consistent with Higgs results and at the same time that
predictions agree with data from neutrino oscillations effectively constrains the parameter space of \bRPV. 
A corresponding fitting routine is implemented in the SPheno code~\cite{Porod:2003um} fulfilling these experimental
constraints in determining \bRPV couplings, spectra and decays. 
Note also that in general all the resulting \bRPV parameters are nonvanishing and are not related in
a trivial way. 
As discussed, for example, in~\cite{DeCampos:2010yu}, one expects strong correlations between neutralino decay
properties measurable at high-energy collider experiments and neutrino mixing angles determined in low-energy
neutrino oscillation experiments, such as
\begin{equation}
\tan^2\theta_{atm} \simeq \frac{BR(\ninoone \to \mu W)}{BR(\ninoone \to \tau W)}.
\label{tantheta23_BRmutau}
\end{equation}
Due to the small size of the \bRPV couplings, the production processes and the SUSY cascade decays are 
usually the same as in corresponding RPC scenarios. 
The fundamental difference in high-energy collision processes arises from decays of the LSP. 
Focusing on prompt LSP decays can lead to two-body decays of a neutralino LSP into gauge boson plus lepton, as described below.

\subsection{\bRPV mSUGRA model}
\label{sec:bRPV_models}
The analyses of~\cite{deCampos:2007bn} and subsequently~\cite{DeCampos:2010yu,deCampos:2012pf} have investigated the corresponding phenomenology and 
expected sensitivities at the LHC. 
Assuming a neutralino LSP being sufficiently heavy, its most relevant two-body decay modes have been discussed:
\begin{itemize}
\item $\ninoone \to \Wboson \tau$
\item $\ninoone \to \Wboson e$
\item $\ninoone \to \Wboson \mu$
\item $\ninoone \to \Zboson \nu$
\item $\ninoone \to h^0 \nu$
\end{itemize}
Notably, for a large part of the parameter space, the decays to $\Wboson \tau$ and $\Wboson \mu$, see 
also~\eqref{tantheta23_BRmutau} tend to be
dominant, requiring consistency with neutrino oscillations. 
The exact magnitude of the individual LSP branching ratios also depend on its couplings, that is, if it corresponds mainly to a bino-, wino- or higgsino-like state.
Ideally, the searches for bRPV SUSY signatures could utilize the subsequent decay products of gauge or Higgs bosons, see, for example,~\cite{DeCampos:2010yu}.   
Reconstructing such bosons accompanying the leptonic partner from LSP decays would allow to reconstruct LSP masses and also indicate its two-body decays to possibly reveal RPV of bilinear type.
Depending on the number of charged leptons in the LSP decays, the phenomenology of final states can be classified as \emph{leptonic}, \emph{semi-leptonic} or \emph{invisible decays}~\cite{deCampos:2007bn}. 
The latter decay mode $\ninoone \to \nu \nu \nu$ would mimic RPC SUSY signal with large \met. 
Moreover it has been emphasized in~\cite{DeCampos:2010yu} that reducing the LSP mass can lead to significantly late LSP decays. 
In the context of mSUGRA the LSP mass is mainly driven by the input parameter $m_{1/2}$, leading to an approximate displacement of decays of 1~mm (in the rest frame of the LSP) for  $m_{1/2}\approx$~300~GeV.

Similar to RPC mSUGRA models, the most relevant production processes are given by: 
\begin{itemize}
\item \gluino-\gluino\ production is most relevant in the region of low $m_{1/2}$; 
\item squark-\gluino\ processes are most significant for low  $m_{1/2}$ and $m_{0}$;
\item contributions from squark-(anti)-squark are most relevant for low  $m_{0}$ and relatively high
$m_{1/2}$;
\item electroweak gaugino-gaugino-based production tends to be dominant for highest input mass scales
of  $m_{0}$ and  $m_{1/2}$.
\end{itemize}
In the minimal supergravity model~\cite{Chamseddine:1982jx,Barbieri:1982eh,Ibanez:1982ee,Hall:1983iz,Ohta:1982wn,Kane:1993td}, the SUSY breaking
sector at the high scale of unification connects to the MSSM at the electroweak scale dominantly through gravitational-strength interactions.
In a minimal form,  one common
mass scale $m_{1/2}$ appears for the three gauginos, one mass scale $m_0$ for all scalars, 
and one coupling $A_0$ for all scalar three-field interactions, so that all gauginos are degenerated and also
all squarks, sleptons, and Higgs-related mass values become degenerate at the unification mass scale.
In addition to these three input parameters, also the ratio of the vevs of the two neutral Higgses,
$\tan\beta$ and the sign of the Higgs mass term, sign$(\mu)$, are necessary to define the mSUGRA model.
After fixing this set of 5 parameters as boundary conditions for mSUGRA, the renormalization group evolution for SUSY-breaking masses and trilinear parameters will finally determine the SUSY mass spectrum at LHC energies. \\
Taking into account the Higgs boson mass observed at 125~\gev, bRPV mSUGRA signal models have been analyzed.
Similar to Higgs-aware signal models of RPC mSUGRA investigated, for example, in~\cite{Aad:2014wea}, the input parameters are chosen as 
$\tan\beta=30$, $A_0=-2 m_0$,  sign($\mu$)=1 with varying values of the mass scales $m_{0}$  and  $m_{1/2}$.
Referring to the same input parameters with respect to RPC mSUGRA also implies the same masses and essentially the same cross-sections in comparison to each RPC-based production process. 
Due to the smallness of \bRPV couplings, other production processes are highly suppressed, so that RPC- and \bRPV-based production processes are almost in one-to-one correspondence in mSUGRA.

\subsection{Results for \bRPV searches}
\label{sec:bRPV_searches}

In the ATLAS SS/3L$_A$ analysis~\cite{Aad:2014pda} the parameter space of \bRPV mSUGRA has been strongly constrained. 
Based on the limits from Figure~\ref{fig:bRPV_prompt_results_SS3L_tau},  values of $m_{1/2}$ are excluded between 200~GeV and 490~GeV at 95\% CL for $m_0$ values below 2.2~\tev. This limit corresponds to a lower bound of approximately 1.3~\tev\ for gluino masses in bRPV mSUGRA. 
Signal models with $m_{1/2} < 200 \gev$ are not considered in this analysis because the lepton acceptance is significantly reduced due to the increased LSP lifetime in that region. The sensitivity is dominated by the signal region SR3b selecting
same-sign or three leptons,  and requiring additionally $\geq 3$~b-\textrm{jets}, $\geq 5$~$N_{\textrm{jets}}$ and $m_{\textrm{eff}}>350$~GeV, respectively.
High sensitivity in particular in signal region SR3b is also a result of the high number of leptons and also of $b$-jets from LSP decays in conjunction with low requirements for missing transverse energy.
It is interesting to note that in SR3b, a 95\% CL upper limit on the (observed) \emph{visible cross-section} at 0.19~fb has been obtained, establishing a model-independent limit. 

The ATLAS $\tau_A$ analysis~\cite{Aad:2014mra} has demonstrated that searching for hadronically decaying $\tau$-leptons in addition to jets, \met and light leptons has a high sensitivity for bRPV at low $m_0$. 
In this part of the parameter space, the number of taus from RPC decays of relatively light staus is high. 
Adding also $\tau$-leptons from \bRPV LSP decays, the number of taus is even more pronounced in this case.
Notably, several SRs, based on $\tau+\mu$, $\tau+e$ and $2\tau$, have been optimized particularly for bRPV.
Performing a statistical combination of these SRs, the 95\%~CL limits on mSUGRA mass parameters
in~\figref{fig:bRPV_prompt_results_SS3L_tau} have been obtained.
As a result from~\cite{Aad:2014mra}, values of $m_{1/2}$ up to 680~\gev\ are excluded
for low $m_{0}$, while the exclusion along the $m_{0}$ axis reaches a maximum of 920~\gev\ for
$m_{1/2}$ = 360~\gev.
For the results in SRs relevant for bRPV searches also limits on visible cross-sections have been
derived, corresponding to upper limits on the observed $\sigma_{\textrm{vis}}$ of 0.52~fb, 0.26~fb and 0.20~fb in
the $\tau+\mu$, $\tau+e$ and $2\tau$ channels, respectively.
Although the expected $\sigma_{\textrm{vis}}$ are the same for $\tau+\mu$
and $\tau+e$, the higher number of events observed in the SR$(\tau+\mu)$ effectively leads to a weaker limit of
0.52~fb with respect to 0.26~fb in SR$(\tau+e)$.

Moreover, the ATLAS search for leptons in SUSY strong production~\cite{Aad:2015mia}, optimized for
RPC models, has also obtained exclusion limits for this \bRPV mSUGRA model. 
Comparing the expected limits to the SS/3L analysis, they are comparable for both analyses in the
range of low $m_0$, whereas, for the highest values of $m_0$, 
the SS/3L analysis reaches slightly stronger expected limits. 
In terms of observed limits, both analyses are actually comparable in the high $m_0$ regime, while~\cite{Aad:2015mia} obtains a stronger limit in the case of low values of $m_0$. 
In particular for $m_0=400$~\gev\, an observed exclusion of
$m_{1/2}\approx$750~\gev\ corresponding to $m_{\gluino}\approx$1.7~\tev\ has been obtained. 
It is interesting to note that, within the SRs investigated in~\cite{Aad:2015mia}, the hard single-lepton channel has the highest sensitivity to the bRPV mSUGRA model.  

\begin{figure}[htbp]
\begin{subfigure}{0.5\textwidth}
\centering
\includegraphics[width=0.9\textwidth]{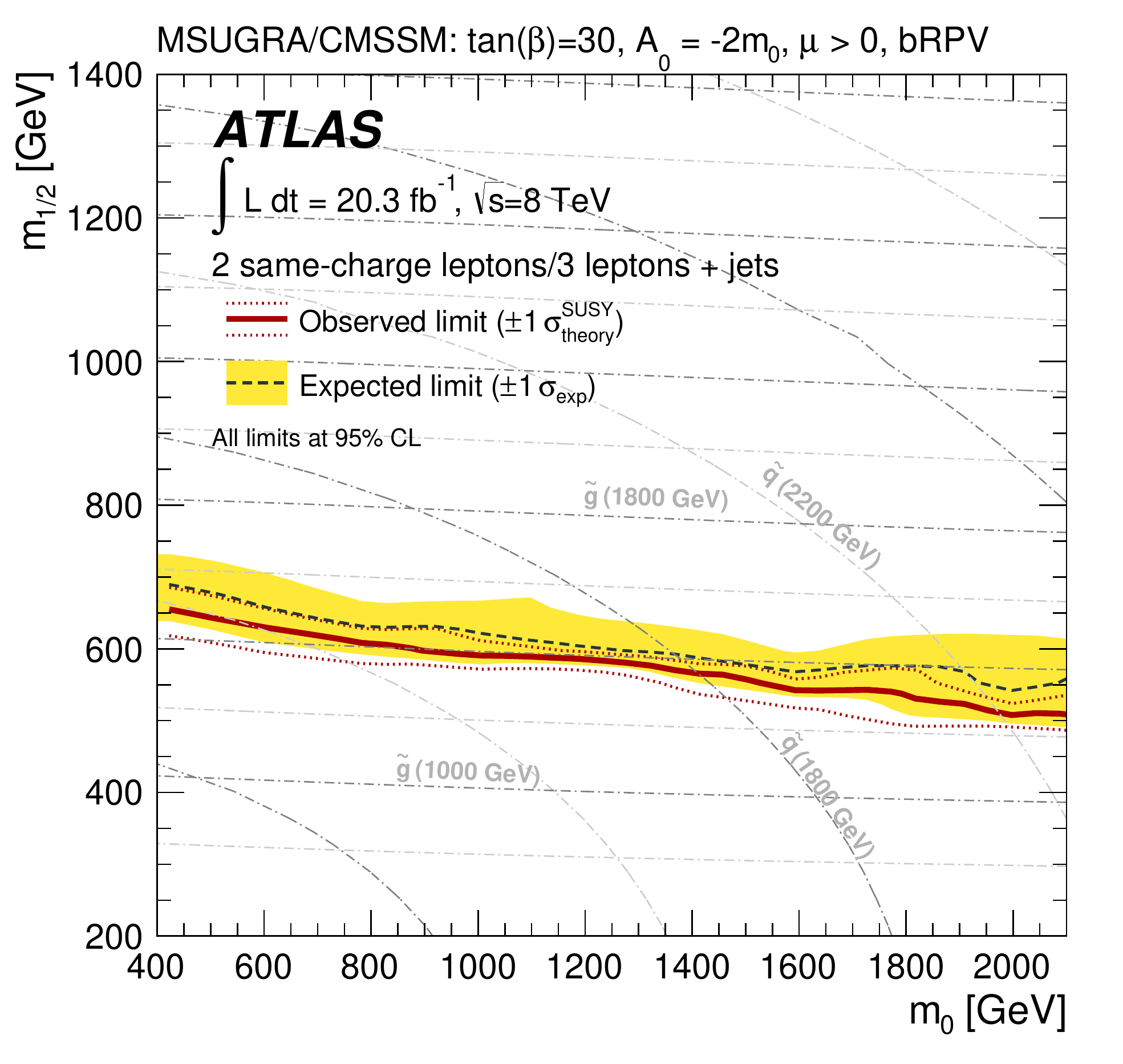}
\caption{bRPV mSUGRA in SS/3L$_A$}
\end{subfigure}
\begin{subfigure}{0.5\textwidth}
\centering
\includegraphics[width=0.9\textwidth]{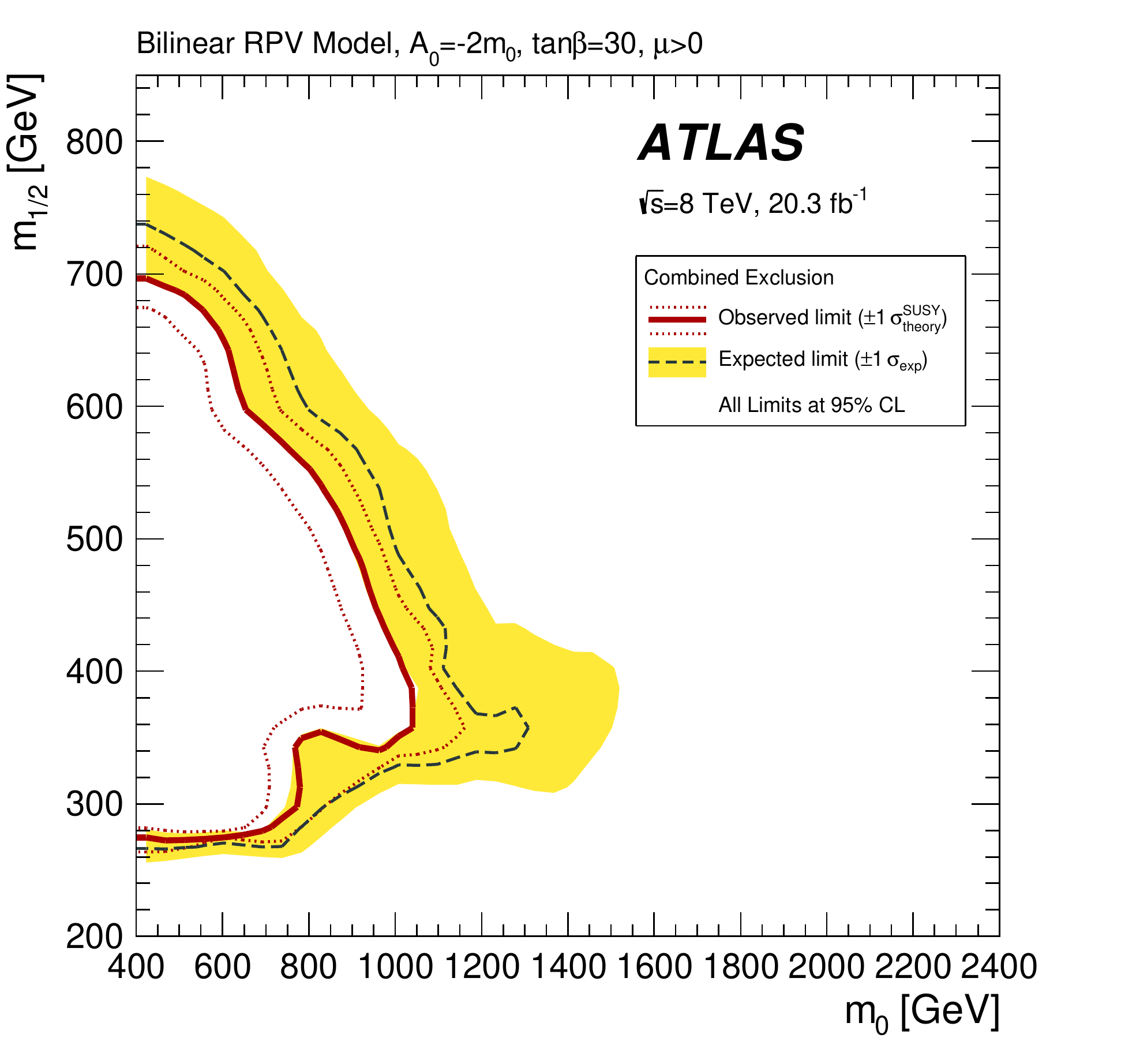}
\caption{bRPV mSUGRA in $\tau_A$}
\end{subfigure}
\caption{Observed and expected exclusion limits for the bRPV mSUGRA model obtained in the
SS/3L ATLAS analysis (left, from~\cite{Aad:2014pda}) and in the $\tau_A$ analysis (right, from
~\cite{Aad:2014mra}).}
\label{fig:bRPV_prompt_results_SS3L_tau}

\end{figure}

\section{\LLE  models}
\label{sec:LLE}
\subsection{\LLE simplified models}
\label{sec:LLEsimplified_neut_LSP}
\noindent
In the RPV simplified models studied in~\cite{Aad:2014iza}, a bino-like \ninoone{} is assumed to decay into two charged leptons and a neutrino via the \lam~term. 
Four event topologies are tested, resulting from different choices for the next-to-lightest SUSY particles (NLSPs):
a chargino (\chinoonepm) NLSP;
slepton NLSPs, referring to mass-degenerate charged sleptons;
sneutrino NLSPs, referring to mass-degenerate sneutrinos;
and a gluino NLSP.
In the slepton case, both the left-handed and right-handed sleptons (L-sleptons and R-sleptons,
respectively) have been considered, as the different production cross-sections for the two cases substantially affect the analysis sensitivity.
The assumed decays of each NLSP choice are described in \tabref{tab:RPVmodeldecay} and illustrated in \figref{fig:RPVgraphs_4L_A}.
The masses of the NLSP and LSP are varied, while other sparticles are assumed to be decoupled.
	
\begin{table}[ht]
\centering
\small{
\begin{tabular}{l c l }
	\hline
	  RPV model NLSP& & Decay   \\ 
	\hline\hline
Chargino      & & $\chinoonepm \rightarrow W^{\pm(*)} ~\ninoone$   \\ 
	  L-slepton     & & $\sleptonL \rightarrow \ell ~\ninoone$    \\ 
	                & & $\stauL \rightarrow \tau ~\ninoone$   \\ 
	  R-slepton     & & $\sleptonR \rightarrow \ell ~\ninoone$    \\ 
	                & & $\stauR \rightarrow \tau ~\ninoone$   \\ 
	  Sneutrino     & & $\tilde{\nu}_l \rightarrow \nu_{\ell} ~\ninoone$    \\ 
	                & & $\tilde{\nu}_\tau \rightarrow \nu_{\tau} ~\ninoone$  \\
	  Gluino        & & $\gluino \rightarrow q \qbar ~\ninoone$     \\ 
	                & & $q \in u,d,s,c$ \\
	\hline\hline
\end{tabular}
}
\caption{Sparticle decays in the SUSY RPV simplified models considered in~\cite{Aad:2014iza}.
	The neutralino LSP is assumed to decay to two charged leptons and a neutrino.
	For the chargino model, the $\Wpm$ from the
	\chinoonepm{} decay may be virtual as indicated by the superscript $(*)$.
\label{tab:RPVmodeldecay}}
\end{table}

\begin{figure}[htbp]
\begin{subfigure}{0.23\textwidth}
\centering
\includegraphics[width=0.8\textwidth]{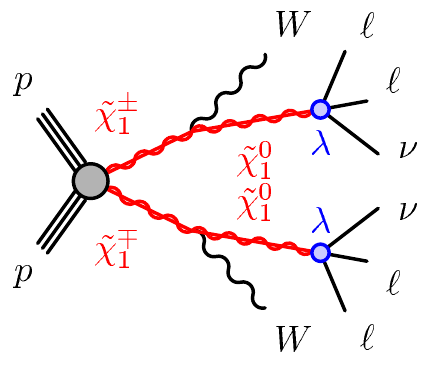}
\caption{~Chargino NLSP}
\label{fig:RPVgraphs_4L_A_chargino-NLSP}
\end{subfigure}
\begin{subfigure}{0.23\textwidth}
\centering
\includegraphics[width=0.8\textwidth]{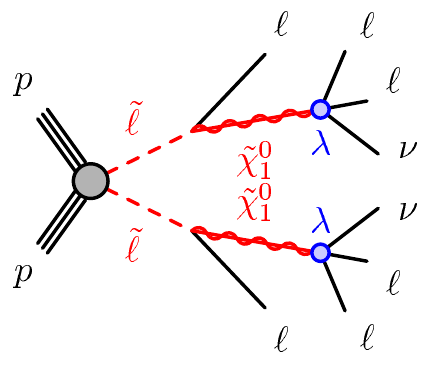}
\caption{~Slepton NLSP}
\label{fig:RPVgraphs_4L_A_RL-slepton-NLSP}
\end{subfigure}
\begin{subfigure}{0.23\textwidth}
\centering
\includegraphics[width=0.8\textwidth]{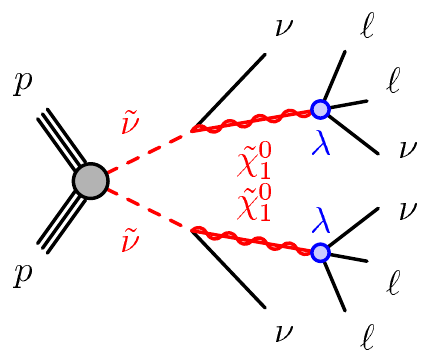}
\caption{~Sneutrino NLSP}
\label{fig:RPVgraphs_4L_A_sneutrino-NLSP}
\end{subfigure}
\begin{subfigure}{0.23\textwidth}
\centering
\includegraphics[width=0.8\textwidth]{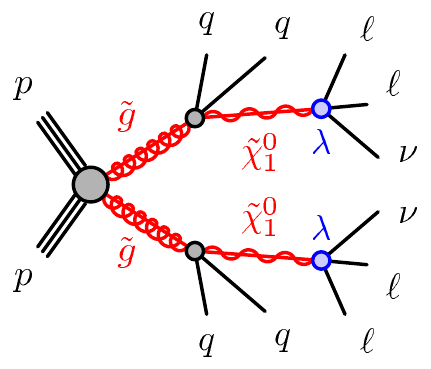}
\caption{~Gluino NLSP}
\label{fig:RPVgraphs_4L_A_gluino-NLSP}
\end{subfigure}
\caption{Representative diagrams for the RPV simplified models based on electroweak or gluino production (from~\cite{Aad:2014iza}).}
\label{fig:RPVgraphs_4L_A}
\end{figure}

In the paper~\cite{Chatrchyan:2013xsw}, \LLE couplings have been investigated in the context of stop-pair production:
The corresponding simplified model assumes stop decays to a top quark and intermediate on- or off-shell bino, $ \stopone \to \ninoone + t$. 
The bino decays to two leptons and a neutrino through the leptonic  RPV interactions,
$ \widetilde{\chi}^{0(*)}_1 \to \ell_i + \nu_j + \ell_k\quad\text{and}\quad\nu_i + \ell_j + \ell_k$, 
where the indices $i, j, k$ refer to those appearing in~\Eqref{eq:LRPV}. 
The stop is assumed to be right-handed and RPV couplings are large enough that all decays are prompt.
Results for the corresponding simplified mass spectra and leptonic RPV couplings $\lam[122]$ or
$\lam[233]$ are investigated in Section~\ref{sec:LLECMS_results}.

\subsection{\LLE RPV results with electroweak or \gluino\ production} 
\label{sec:LLE_results_multi-lepton}
The \LLE simplified models produce events with four leptons in the final state, and thus it is natural to constrain them with the ATLAS search for SUSY in events with four or more charged leptons~\cite{Aad:2014iza}.
Up to two of the leptons may be hadronically decaying taus, and the search was specifically optimised to give good sensitivity across the full range of \LLE-mediated \ninoone{} decays.

In all cases, the observed limit is determined mainly by the production cross-section
of the signal process, with stronger constraints on models where \lam[121] or \lam[122] dominate, and less stringent limits for tau-rich decays via \lam[133] or \lam[233].
Limits on models with different combinations of \lam{} parameters can generically be expected to lie between these extremes.
The limits are in many cases nearly insensitive to the \ninoone{} mass, except where the \ninoone{} is
significantly less massive than the NLSP as 
inferred from~\figref{fig::ExpLimitWinoGluinoSleptonSneutrino_4L_A}.
Where the NLSP$\to$LSP cascade may also produce leptons, the observed limit may also become weaker as $m_{\ninoone}$ approaches the NLSP mass, and the cascade product momenta decrease considerably.

When the mass of the \ninoone{} LSP is at least as large as 20\% of the NLSP mass, and assuming tau-rich LSP decays,
 lower limits can be placed on sparticle masses, excluding
gluinos with masses less than $950\gev$; wino-like charginos with masses less than $450\gev$; and L(R)-sleptons with masses less than
300 $(240)\gev$.
If instead the LSP decays only to electrons and muons,
the equivalent limits are approximately $1350\gev$ for gluinos, $750\gev$ for charginos, 490 $(410)\gev$ for L(R)-sleptons, and a 
lower limit of $400\gev$ can also be placed on sneutrino masses.
These results significantly improve upon previous searches at the LHC, where gluino masses of up to $1\tev$~\cite{Chatrchyan:2012mea} and chargino masses of up to $540\gev$~\cite{ATLAS:2012kr} were excluded.

The model-independent limits on $\sigma_{\mathrm{vis}}$ for RPV-related SRs all lie below 0.5~fb: 
In signal regions requiring at least three light leptons, the observed  95\% CL upper limits on the \emph{visible cross-sections} are below 0.2~fb.
\footnote{It is interesting to note that the ATLAS search for multi-leptons based on 7~\tev\ data~\cite{ATLAS:2012kr} has
obtained limits of approximately 1~fb on $\sigma_{\mathrm{vis}}$ considering also four-body decays of a
stau LSP as motivated, for example,  by~\cite{Desch:2010gi}.} 

\begin{figure}[htbp]
\begin{subfigure}{0.5\textwidth}
\centering
\includegraphics[width=0.9\textwidth]{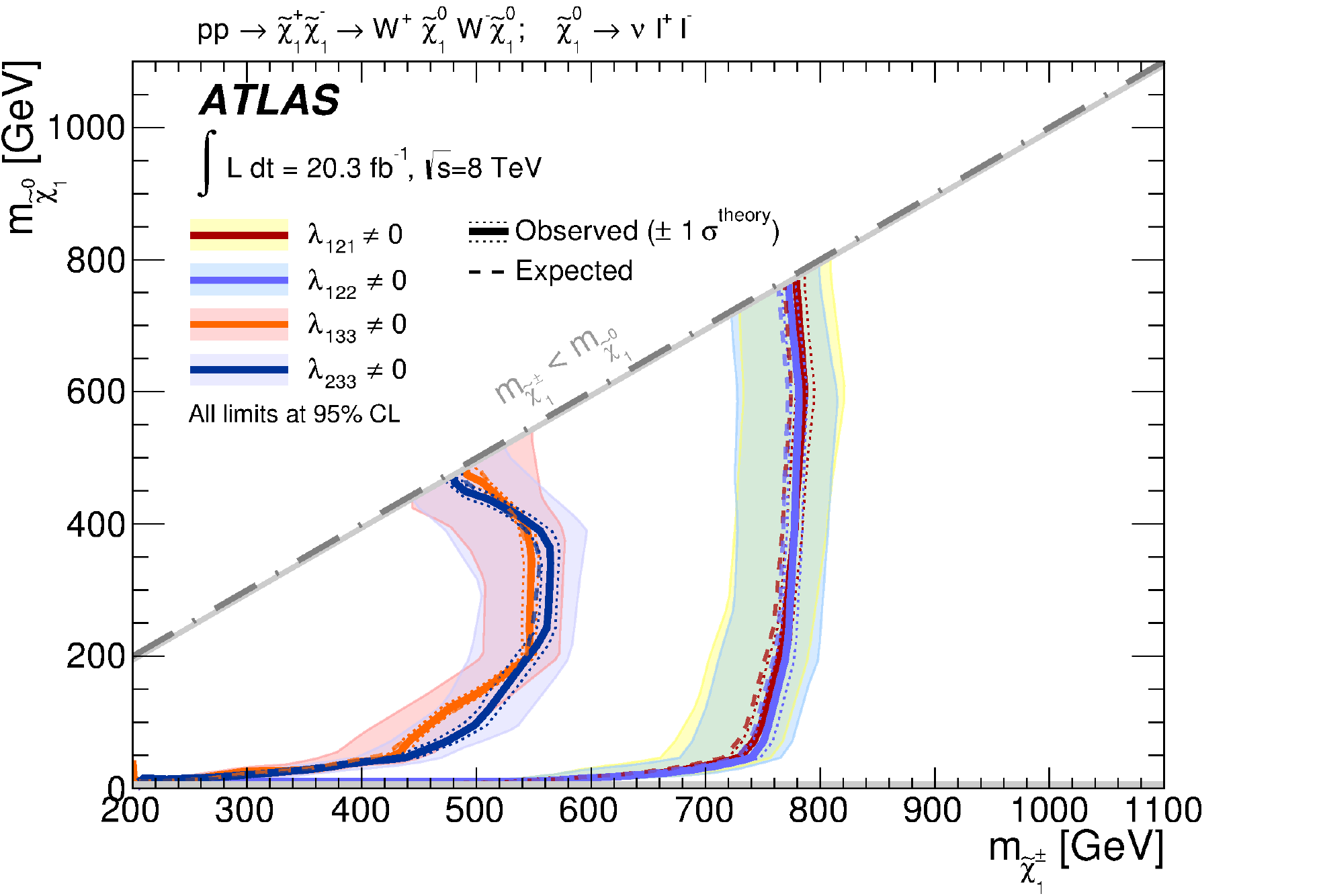}
\caption{~Chargino NLSP}
\end{subfigure}
\begin{subfigure}{0.5\textwidth}
\centering
\includegraphics[width=0.9\textwidth]{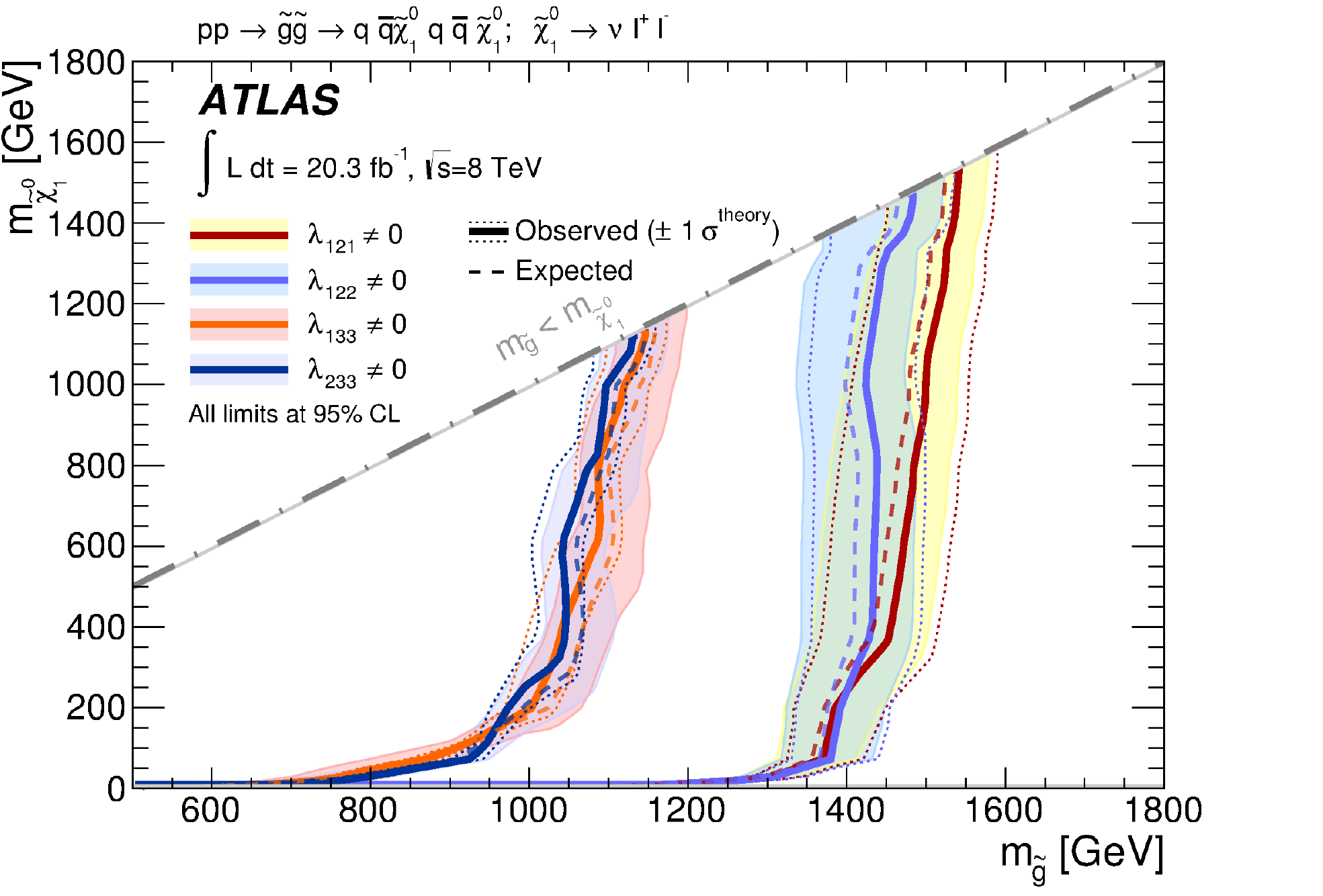}
\caption{~Gluino NLSP}
\end{subfigure}
\begin{subfigure}{0.5\textwidth}
\centering
\includegraphics[width=0.9\textwidth]{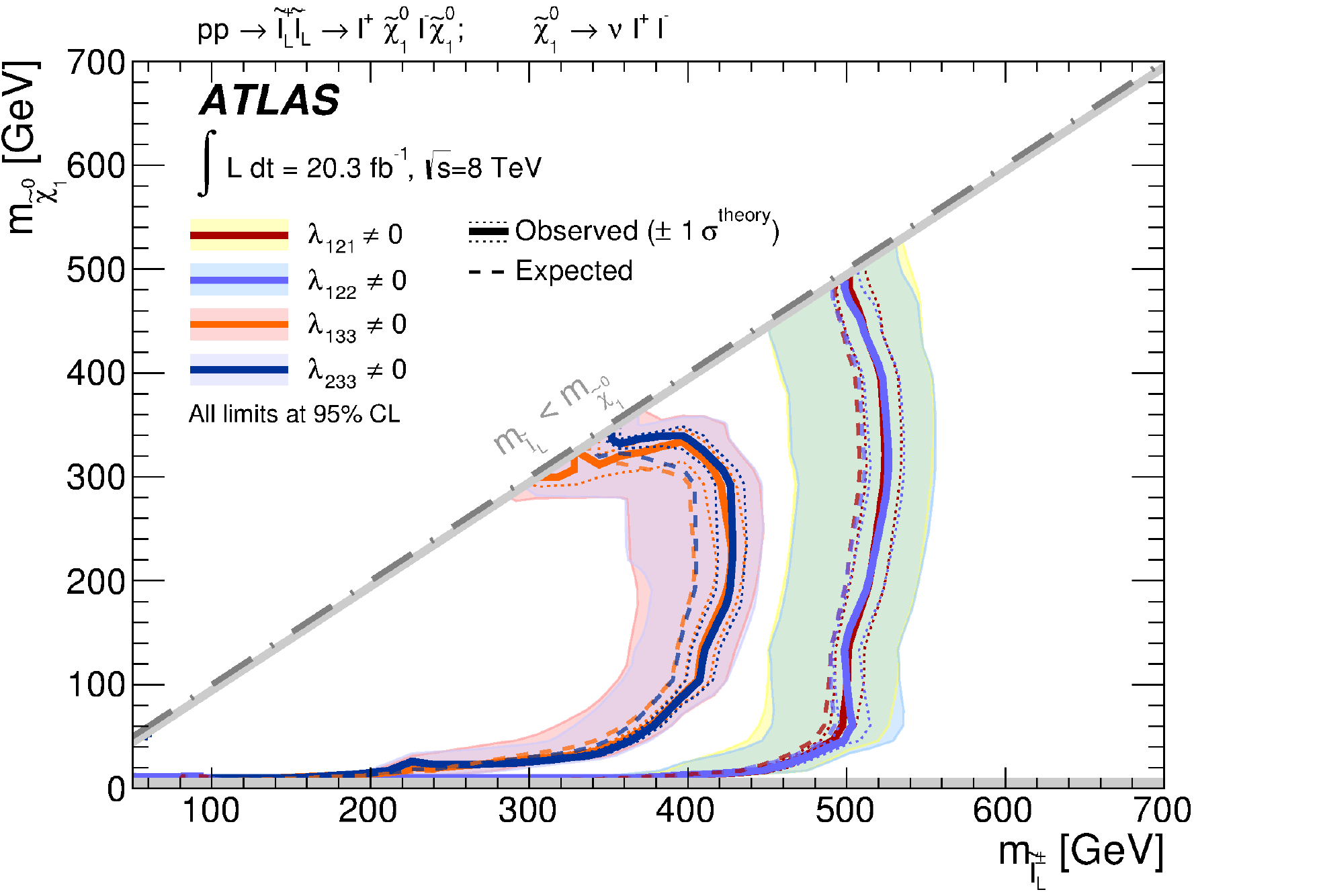}
\caption{~L-slepton NLSP}
\end{subfigure}
\begin{subfigure}{0.5\textwidth}
\centering
\includegraphics[width=0.9\textwidth]{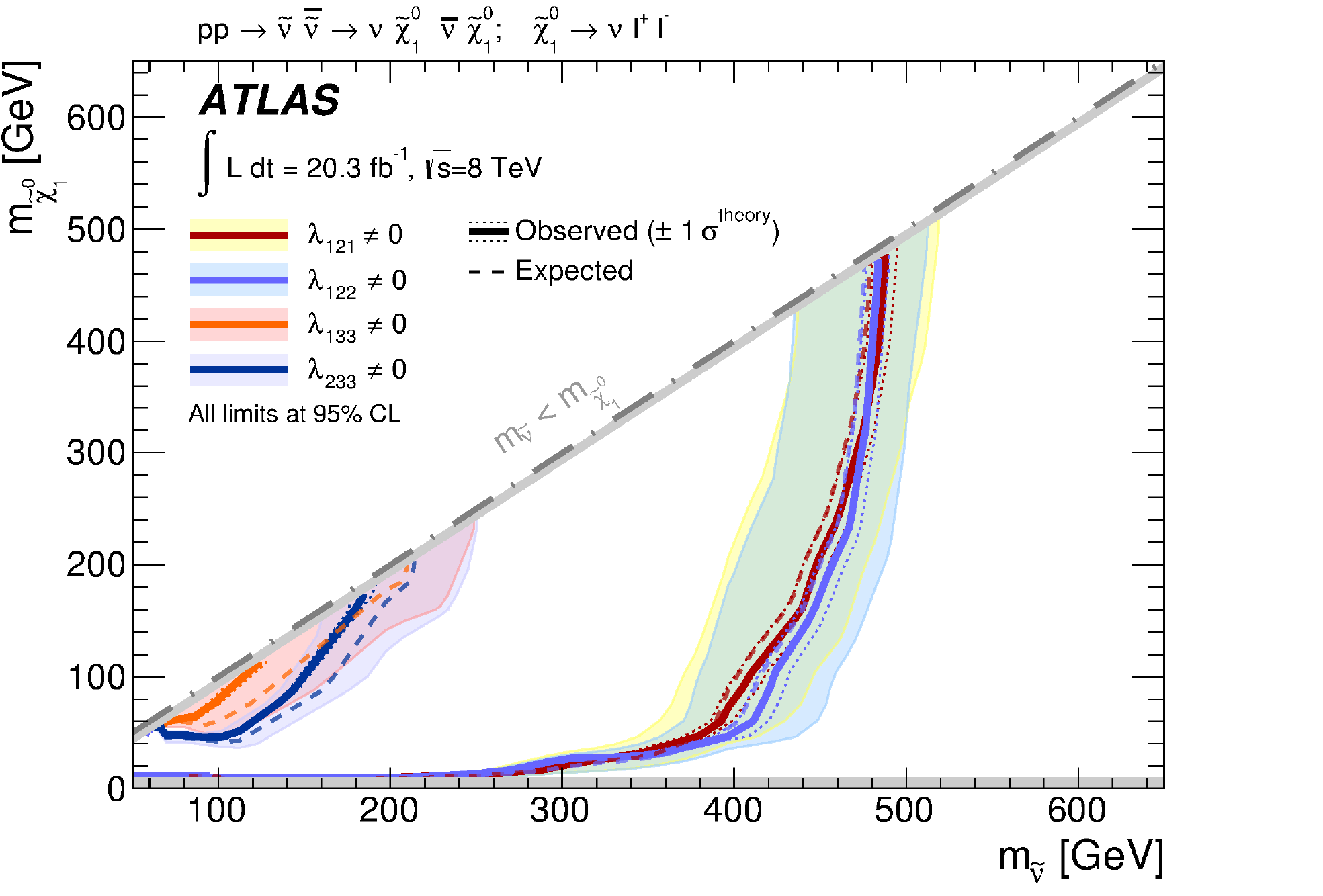}
\caption{~Sneutrino NLSP}
\end{subfigure}
\caption{Observed and expected 95\% CL exclusion limit contours for the \LLE RPV 
(a) chargino NLSP,  
(b) gluino NLSP,
(c) left-handed slepton NLSP and
(d) sneutrino NLSP simplified models
 (from~\cite{Aad:2014iza}).
}
\label{fig::ExpLimitWinoGluinoSleptonSneutrino_4L_A}
\end{figure}

\subsection{\LLE RPV results for \stopone\ production}
\label{sec:LLECMS_results}

 The limits obtained in~\cite{Chatrchyan:2013xsw} are mostly independent of the bino mass, 
leading to an exclusion of models with the stop mass below 1020~\gev\ when $\lam[122]$ is non-zero, and below 820~\gev\ when $\lam[233]$ is non-zero. These limits are shown in Fig. \ref{fig:stopRPV_LLE_3Lb_C}. 
There is a change in kinematics at the line  $ m_{\ninoone} = m_{\stopone} - m_{t}$, below which the stop decay is two-body, while above it is a four-body decay. 
Near this line, the $\ninoone$ and top are produced almost at rest, which results in low-momentum leptons, corresponding to reduced acceptance. 
This loss of acceptance is more visible in the $\lam[233] \neq 0$ case and causes the loss of sensitivity near the line at $m_{\ninoone} = 800$~\gev. 
The analysis~\cite{Chatrchyan:2013xsw} has also explained that this effect is more pronounced in the observed limit because the data has a larger statistical uncertainty in the relevant signal regions than the simulated signal samples.

\begin{figure}[htbp]
\begin{subfigure}{0.5\textwidth}
\centering
\includegraphics[width=0.9\textwidth]{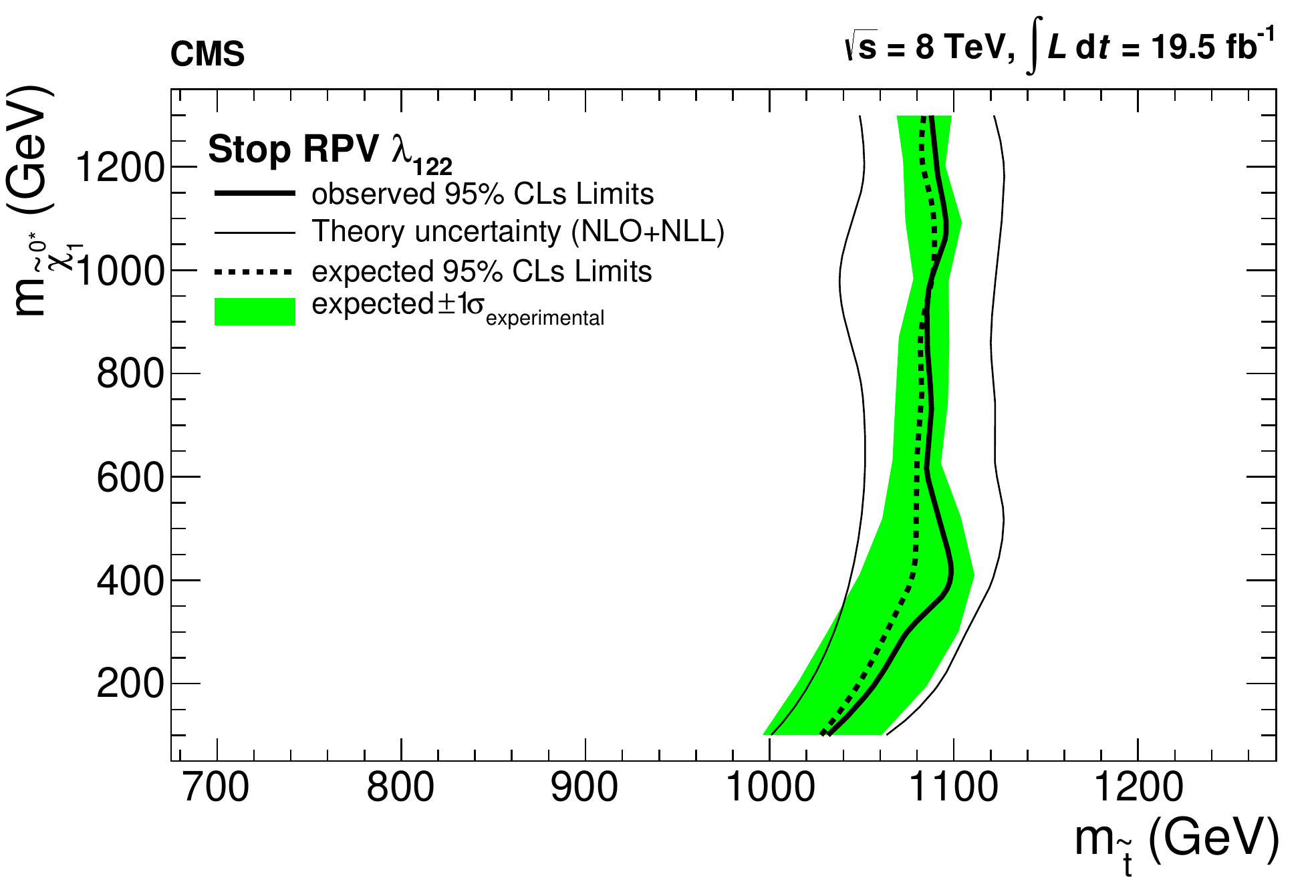}
\caption{~\lam[122]}
\end{subfigure}
\begin{subfigure}{0.5\textwidth}
\centering
\includegraphics[width=0.9\textwidth]{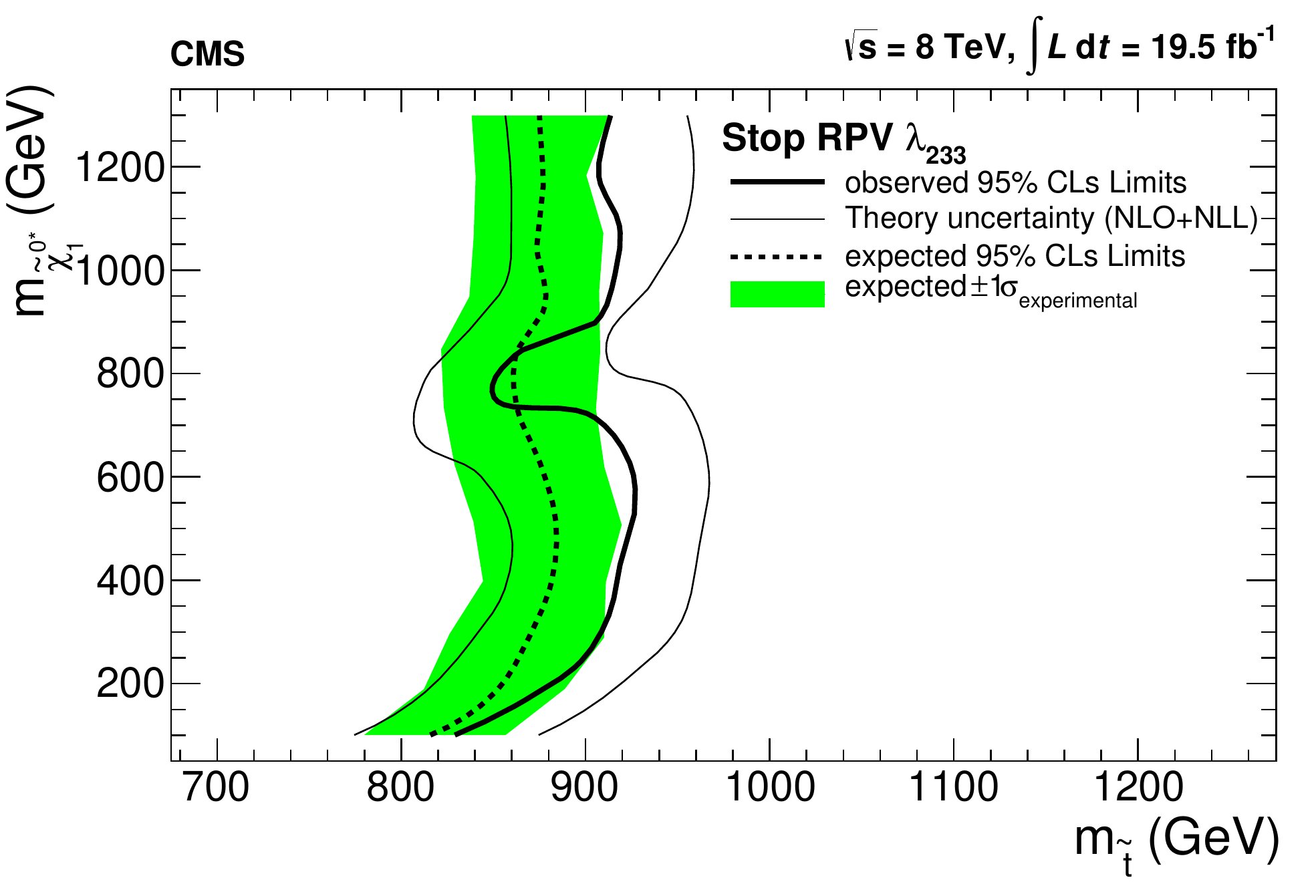}
\caption{~\lam[233]}
\end{subfigure}
   \caption{The 95\% confidence level limits in the stop and bino mass plane for models with RPV couplings $\lam[122]$ and $\lam[233]$. The region to the left of the curve is excluded (from~\cite{Chatrchyan:2013xsw}). 
      }
   \label{fig:stopRPV_LLE_3Lb_C}
\end{figure}

\section{\LQD  models}
\label{sec:LQD}

\subsection{Simplified \LQD models for \stopone\ production}
\label{sec:LQDmodels_3LB-C}

In addition to the simplified model for stop pair production introduced in the previous Section~\ref{sec:LLEsimplified_neut_LSP},
RPV decays via $\lamp[233]$ are also considered in the simplified model of~\cite{Chatrchyan:2013xsw}.
The same assumptions on stop decays to a top quark and intermediate on- or off-shell bino, $ \stopone \to \ninoone + t$ are made, only the \LQD-related decay of $\ninoone$ to one lepton and two quarks leads to different final states
in comparison to LSP decays via \LLE as already discussed in the previous section.  
A possible signal process is illustrated in the Feynman diagram, Figure~\ref{fig:diag_stop_LQD233_3Lb_C}.
Due to the high number of $W$-bosons indicated in the final states of that process, also a relatively large number of charged (light) leptons can be expected, in conjunction with many $b$-jets. 

\begin{figure}[htbp]
\centering
\includegraphics[width=0.4\textwidth]{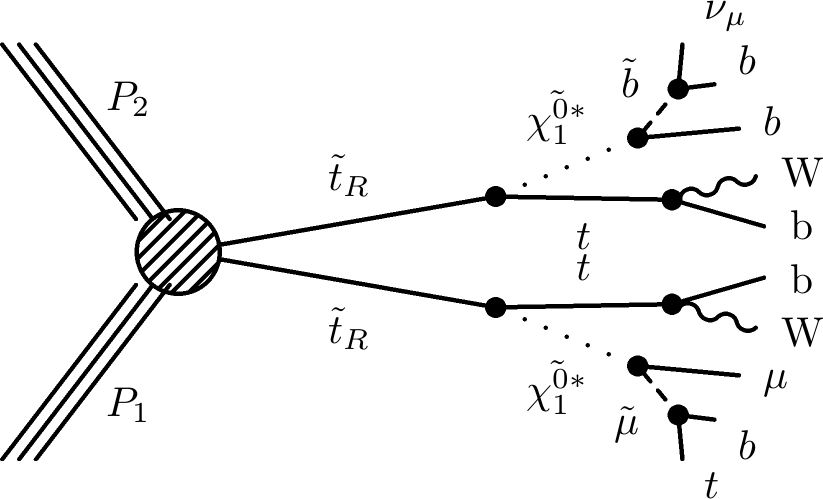}
   \caption{Signal process for stop-pair production with LSP decays mediated via $\lamp[233]$ (from~\cite{Chatrchyan:2013xsw}).
      }
   \label{fig:diag_stop_LQD233_3Lb_C}
\end{figure}

In another simplified model investigated in~\cite{Khachatryan:2014ura}, two different
decay channels of directly produced top squarks are considered.
In the first case the two-body lepton number violating decay
$\stopone \to\tau b$ via the coupling constant
$\lamp[333]$ is investigated; see also~\cite{Barbier:2004ez} for related phenomenological studies.

In the second part of the search the focus is on a scenario
in which the dominant RPC decay of the top squark is $\stopone \to\chinoonepm b$.
This requires the mass splitting between the top squark and the chargino
to be less than the mass of the top quark, so it is chosen to be 100~\gev.
The chargino is assumed to be a pure higgsino
and to be nearly degenerate in mass with the neutralino.
In particular, the decay 
$\chinoonepm\to \tilde{\nu}\tau^{\pm} \to q q\tau^{\pm}$
via an intermediate $\tau$-sneutrino is considered.
This RPV decay of the sneutrino is possible via the \LQD-type coupling $\lamp[3jk]$, where the cases $j, k = 1, 2$ are taken into account.

\subsection{Results  for \stopone\ production} 
\label{sec:LQDCMS_results}

The analysis~\cite{Chatrchyan:2013xsw} has probed regions in the mass plane of neutralino vs. stop masses assuming pair production of \stopone\ and non-vanishing  $\lamp[233]$.
As discussed before, in that analysis several different kinematic regions in the mass plane are relevant also in the final  results of~\figref{fig:stopRPV_LQD_3Lb_C}.
The most significant effect is when the decay $ \widetilde{\chi}^{0}_1 \rightarrow \mu + t + b$ is
suppressed, reducing the number of leptons in the final state. The different regions where this effect is
pronounced primarily lead to the shape of the exclusion for $\lamp[233]$. 
As a result, stop masses up to approximately 800~GeV can be excluded in this model.

\begin{figure}[htbp]
\centering
\includegraphics[width=0.7\textwidth]{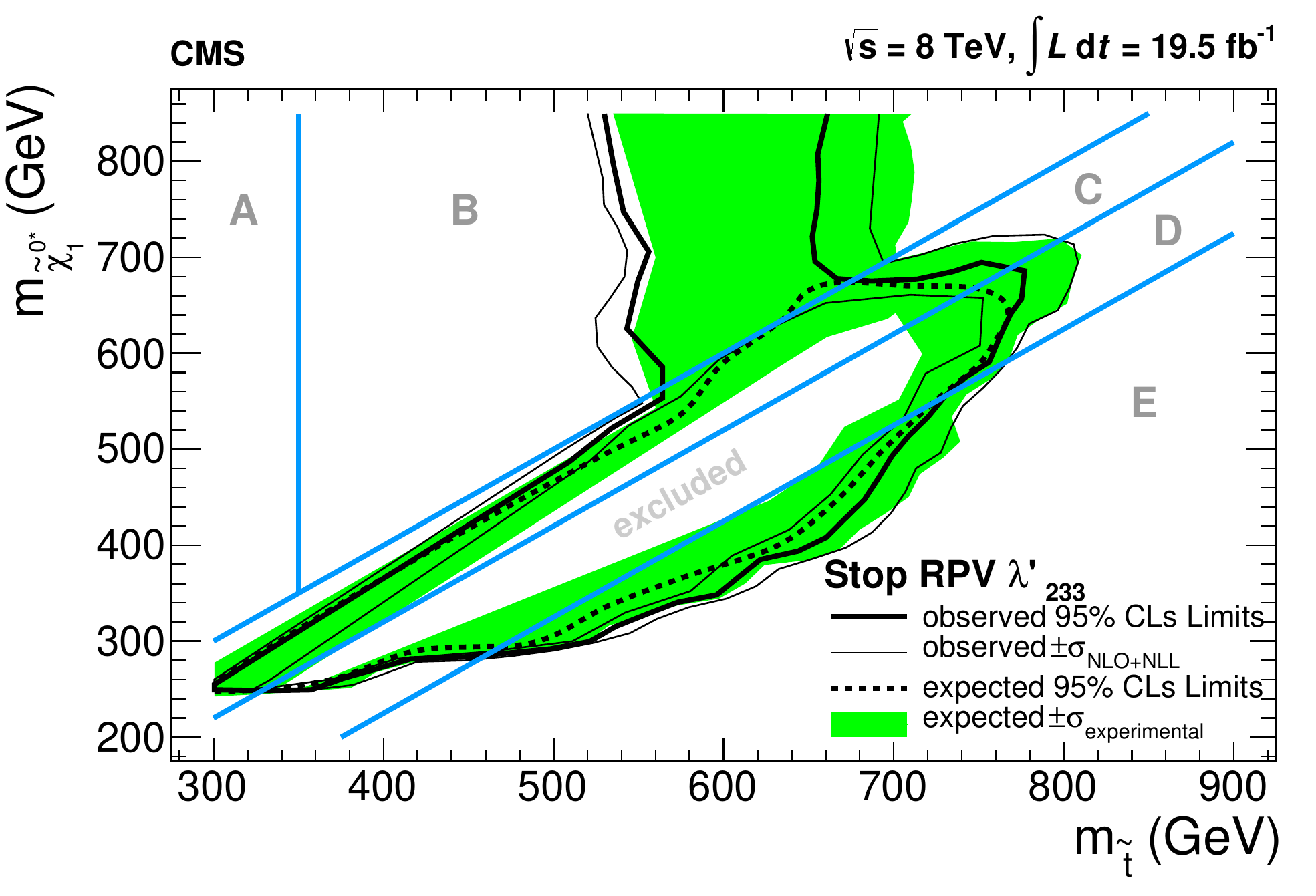}
   \caption{The 95\% CL limits in the stop and bino mass plane for stop-pair production and RPV coupling $\lamp[233]$ (from~\cite{Chatrchyan:2013xsw}). 
The different kinematic regions, A, B, C, D and E are defined in~\cite{Chatrchyan:2013xsw}.
      }
   \label{fig:stopRPV_LQD_3Lb_C}
\end{figure}

In the search for $b$-jets and $\tau$-leptons from CMS~\cite{Khachatryan:2014ura}, constraints for the
masses of pair-produced stops have been derived.
An upper bound at 95\% confidence level is set on $\sigma \mathcal{B}^2$, where
$\sigma$ is the cross-section for pair production of top squarks
and $\mathcal{B}$ is the branching fraction
for the top squark decay to a $\chinoonepm$ and a bottom quark, with a
subsequent decay of the chargino via
$\chinoonepm\to \tilde{\nu}\tau^{\pm} \to q q\tau^{\pm}$.
Expected  and observed upper limits on $\sigma \mathcal{B}^2$
as a function of the stop mass are shown in
Fig.~\ref{fig:LQD_333_LQD_3kj_taub_C} for the top squark search from~\cite{Khachatryan:2014ura}.
As a result, top squarks undergoing a chargino-mediated decay involving the coupling $\lambda^{\prime}_{3jk}$
with masses in the range 200--580~GeV are excluded, in agreement
with the expected exclusion limit in the range 200--590~GeV. In the derivation of these upper limits $\mathcal{B}=100\%$ is assumed.

Since the other simplified model investigated in~\cite{Khachatryan:2014ura} leads to direct decays of stops after $\stopone$ pair-production, the underlying stop mass essentially determines the results.
The limits corresponding to top squarks decaying directly through the coupling $\lambda^{\prime}_{333}$
exclude masses of $\stopone$ below 740~\gev, in agreement with the expected limit at 750~\gev.

\begin{figure}[htbp]
\begin{subfigure}{0.5\textwidth}
\centering
\includegraphics[width=0.9\textwidth]{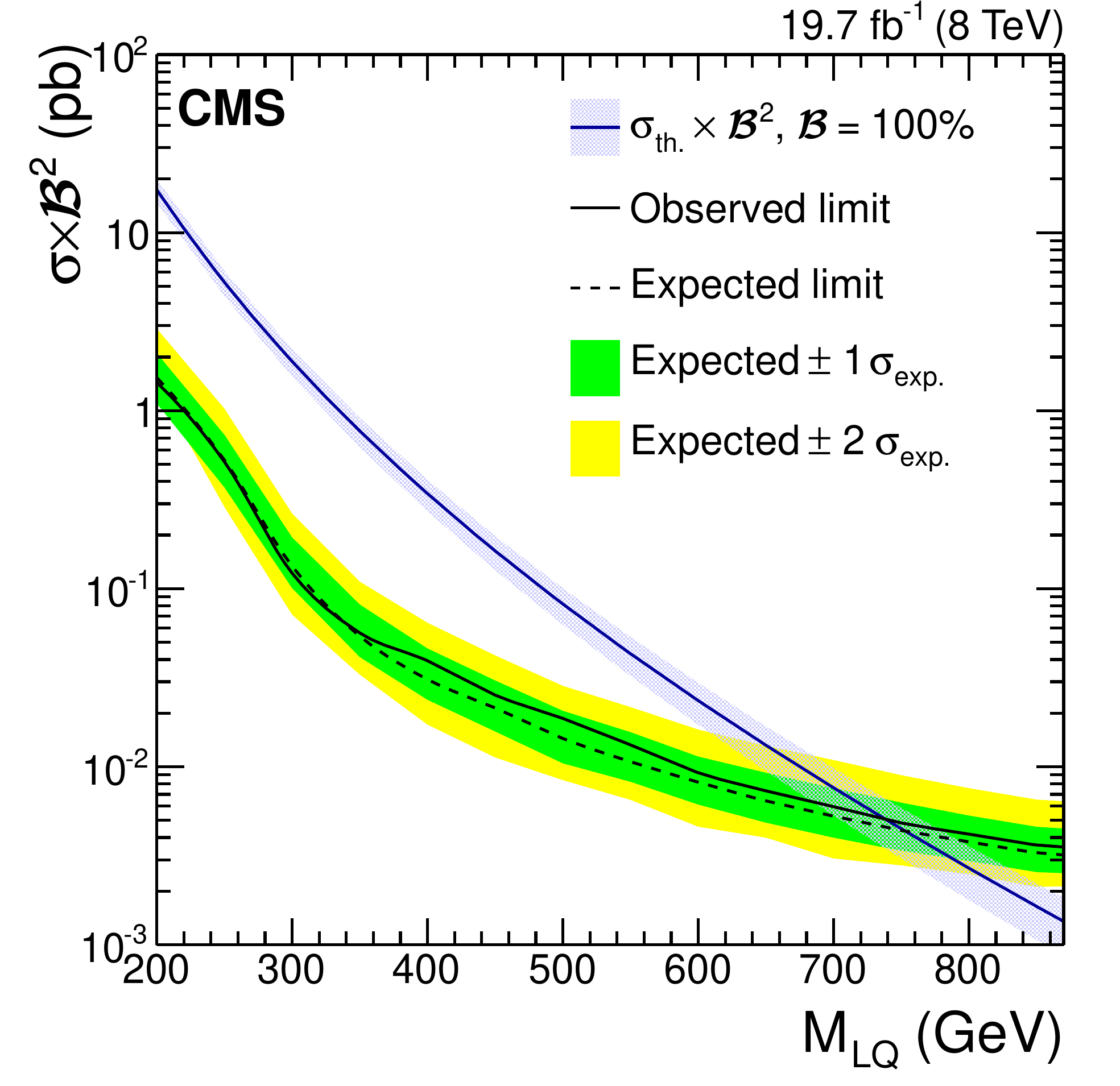}
\caption{~\lamp[333]}
\end{subfigure}
\begin{subfigure}{0.5\textwidth}
\centering
\includegraphics[width=0.9\textwidth]{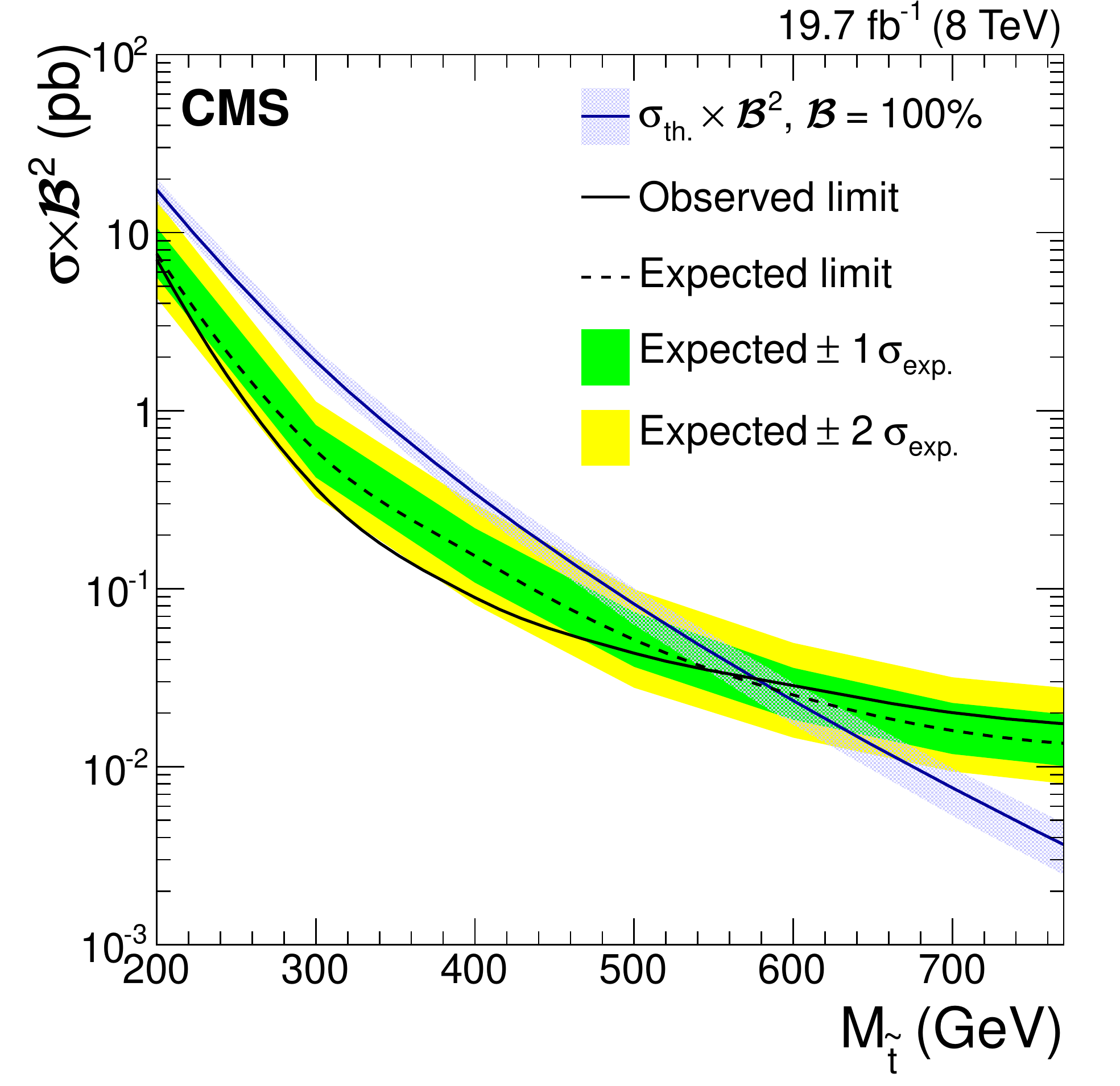}
\caption{~\lamp[3kj]}
\end{subfigure}
   \caption{The expected (observed) combined upper limits on the third-generation LQ pair production
   cross-section $\sigma$ times the square of the branching fraction, $\mathcal{B}^2$. These limits also apply to top squarks decaying directly via the coupling $\lamp[333]$ (left).
The limits (right) apply to top squarks with a chargino-mediated decay through the coupling $\lamp[3kj]$ (from~\cite{Khachatryan:2014ura}). 
      }
   \label{fig:LQD_333_LQD_3kj_taub_C}
\end{figure}

\section{Resonance production and decay}
\label{sec:Resonance}
If \LQD couplings are present at hadron colliders, resonant production of sleptons is possible.
The decay products from such a slepton resonance can also depend on additional RPV couplings.
If also \LLE couplings occur, then leptonic final states can be investigated, whereas jets are expected if only \LQD terms are assumed.
In order to allow for considerable production rates and also for significant decay rates into charged
leptons, $\tau$-sneutrinos emerge as candidates for resonance 
searches.
\footnote{It is interesting to note that also a search for RPV resonances from second generation sleptons has been performed by CMS at
7~TeV~\cite{CMS:2013lda}. Assuming the coupling $\lamp[211]$, the search investigated two same-sign muons and at least two jets in the final state.}
In the case of $\tilde{\nu}_\tau$, the corresponding bounds for its coupling of type \LQD and also \LLE are relatively weak. 
Therefore the analysis~\cite{Aad:2015pfa} searching for resonances using leptonic final states has focused on $\tau$-sneutrinos, as described in more detail below.

\subsection{Resonance via tau sneutrino}
\label{sec:Resonance_tau sneutrino}

As illustrated in~\figref{fig:diag_Resonance}, a $\tau$ sneutrino ($\tilde{\nu}_\tau$) may be produced in $pp$ collisions by $d\bar{d}$
annihilation and subsequently decay to $e\mu$, $e\tau$, or $\mu\tau$.
Although only $\tilde{\nu}_{\tau}$ is considered in~\cite{Aad:2015pfa} to facilitate comparisons with 
previous searches performed at the Tevatron, the results of this analysis in principle apply to any 
sneutrino flavor. 

\begin{figure}
\centering
\includegraphics[width=0.3\textwidth]{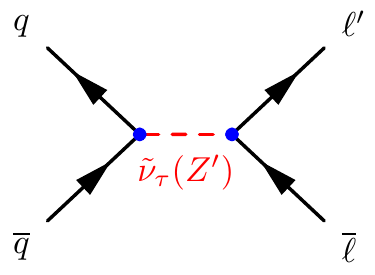}
   \caption{Feynman diagram illustrating resonant production and decay of $\tilde{\nu}_{\tau}$. 
 $l_i$ and $l_j$ can be $e$, $\mu$ or $\tau$, where $i\neq j$.
The blue dots indicate RPV couplings $\lamp[311]$ and $\lam[i3j]$ ($i\neq j$), respectively (from~\cite{Aad:2015pfa}).
      }
   \label{fig:diag_Resonance}
\end{figure}

\subsection{Resonance searches}
\label{sec:Resonance_searches}

Expected and observed upper limits are set as a function of  $\tilde{\nu}_{\tau}$ mass. 
The likelihood of observing the number of events in data as a function of the expected number of signal 
and background events is constructed from a Poisson distribution for each bin in the $\tilde{\nu}_{\tau}$ mass. 
Signal cross-sections are calculated to next-to-leading order for $\tilde{\nu}_{\tau}$.

\noindent
Figure~\ref{fig:Resonance_obsexcl_sneut-tau_mass} 
shows the observed and expected cross-section times branching ratio 
limits as a function of the $\tilde{\nu}_{\tau}$ mass.  
For a $\tilde{\nu}_{\tau}$ mass of 1~TeV, the observed limits on the production cross-section 
times branching ratio are 0.5 fb, 2.7 fb, and 9.1 fb for the $e\mu$, $e\tau$ and $\mu\tau$ channels, 
respectively.
Theoretical predictions of cross-section times branching ratio
are also shown, 
assuming $\lambda'_{311}=0.11$ and $\lambda_{i3k}=0.07$ ($i\neq k$) for 
the $\tilde{\nu}_{\tau}$, consistent with benchmark 
couplings used in previous searches. 
For these benchmark couplings, the lower limits on the $\tilde{\nu}_{\tau}$  mass 
are 2.0 TeV, 1.7 TeV, and 1.7 TeV for the $e \mu$, $e \tau$ 
and $\mu \tau$ channels, respectively.

\noindent
These results considerably extend previous constraints from the Tevatron and LHC experiments. 
Based on similar assumptions for RPV couplings, that is,  $\lambda'_{311}=0.10$ and $\lambda_{i3k}=0.05$ ($i\neq k$), the CDF experiment~\cite{Aaltonen:2010fv} has obtained lower limits for 
$\tau$-sneutrino masses at
                558~\gev, 442~\gev\ and 441~\gev\
  for the $e \mu$, $e \tau$  and $\mu \tau$ channels, respectively.

\begin{figure}
\centering
\includegraphics[width=0.9\textwidth]{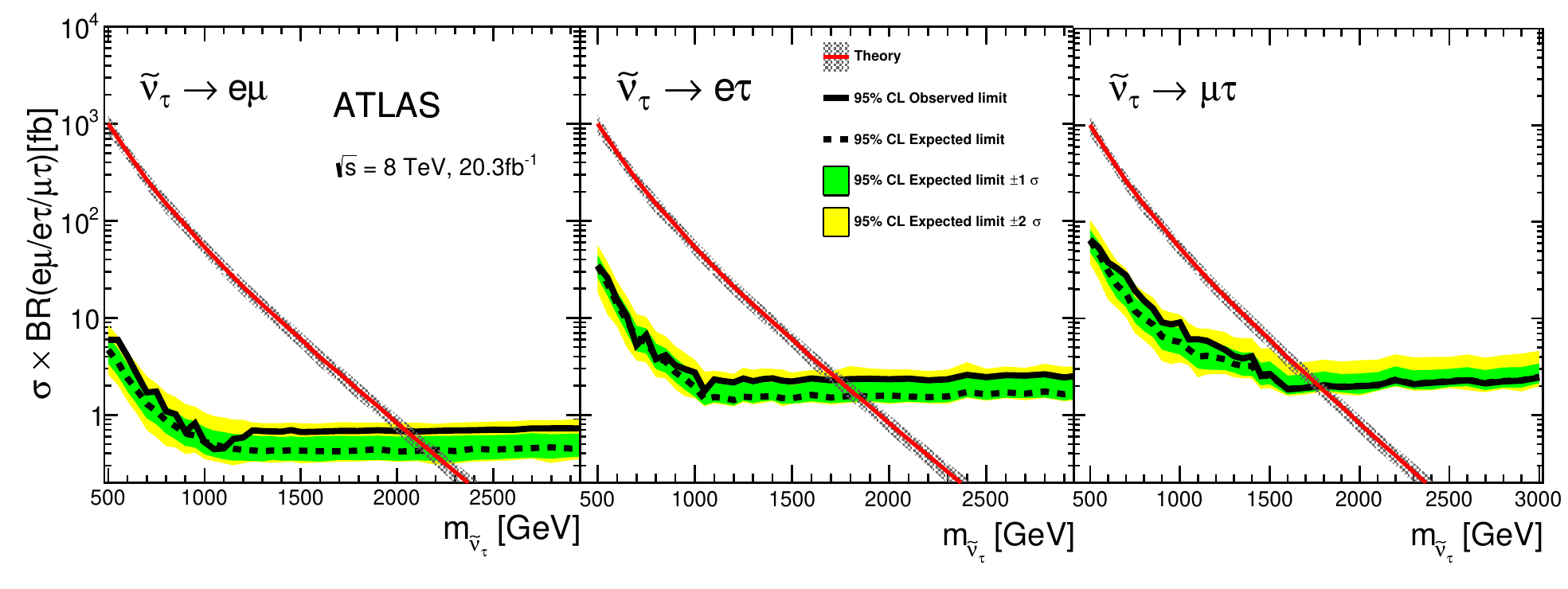}
   \caption{The 95\% CL limits on cross-section times branching ratio as a function of $\tilde{\nu}_\tau$ 
mass for $e\mu$ (left), $e\tau$ (middle), and $\mu\tau$ (right)  (from~\cite{Aad:2015pfa}).  
Theory curves correspond to $\lambda'_{311}=0.11$ and $\lambda_{i3k}=0.07$ for 
$\tilde{\nu}_{\tau}$. 
      }
   \label{fig:Resonance_obsexcl_sneut-tau_mass}
\end{figure}

\section{UDD  models}
\label{sec:UDD}
In this section results from searches for signatures from \UDD couplings are presented. 
Both ATLAS and CMS have investigated several different topologies as motivated by a number of simplified models. 

\subsection{\gluino\ production with multi-jets at ATLAS}
\label{sec:UDDmulti-jet_models_ATLAS}
Pair-produced massive new particles decay directly to a total of six quarks, as well as cascade decays with at least ten quarks, are considered in the design of the analysis~\cite{Aad:2015lea}. 
Three-body decays of the type shown in \figref{fig:UDD_Feynman_diagrams_multi-jets} are given by effective RPV vertices allowed by the baryon-number-violating \lamdp\ couplings with off-shell squark propagators. 
For both models, all squark masses are set to 5~TeV and thus gluinos decay directly to three quarks or to two quarks and a neutralino through standard RPC couplings. In the ten-quark cascade decay model, the neutralinos each decay to three quarks via an off-shell squark and the RPV \UDD decay vertex with coupling \lamdp. In this model, the neutralino is the LSP. 

 All possible $\lamdp[]$ flavor combinations are allowed to proceed with equal probability. The analysis maintains approximately equal sensitivity to all flavor modes.  All samples are produced assuming that the gluino and neutralino widths are narrow and that their decays are prompt.

  \begin{figure}[htbp]
    \centering
    \includegraphics[width=0.3\columnwidth]{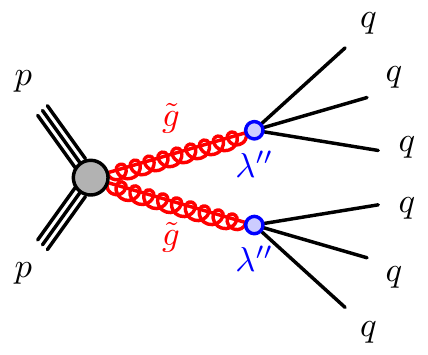}
    \includegraphics[width=0.3\columnwidth]{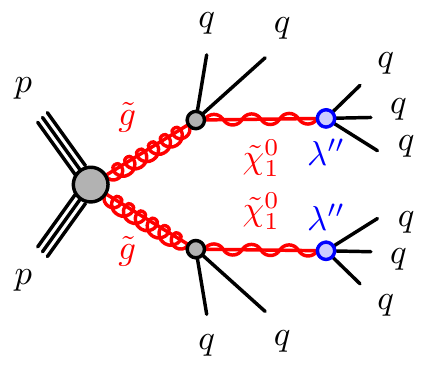}
    \caption{Diagrams for multi-jet processes considered for the analysis~\cite{Aad:2015lea}. The gray shaded circles represent effective vertices that include off-shell propagators, and the blue shaded circles represent effective RPV vertices allowed by \UDD \lamdp\ couplings with off-shell propagators (from~\cite{Aad:2015lea}).}
    \label{fig:UDD_Feynman_diagrams_multi-jets}
  \end{figure}

It is interesting to compare limits based on different assumptions for the branching ratios into heavy flavor
jets. \figref{fig:mass_BR-bjets-multij-A} illustrates the variation for the observed mass limit when the decays into $b$-jets are absent or assumed at 100 percent, respectively.

\begin{figure}[htbp]
\begin{subfigure} {0.5\textwidth}
\centering
    \includegraphics[width=0.9\textwidth]{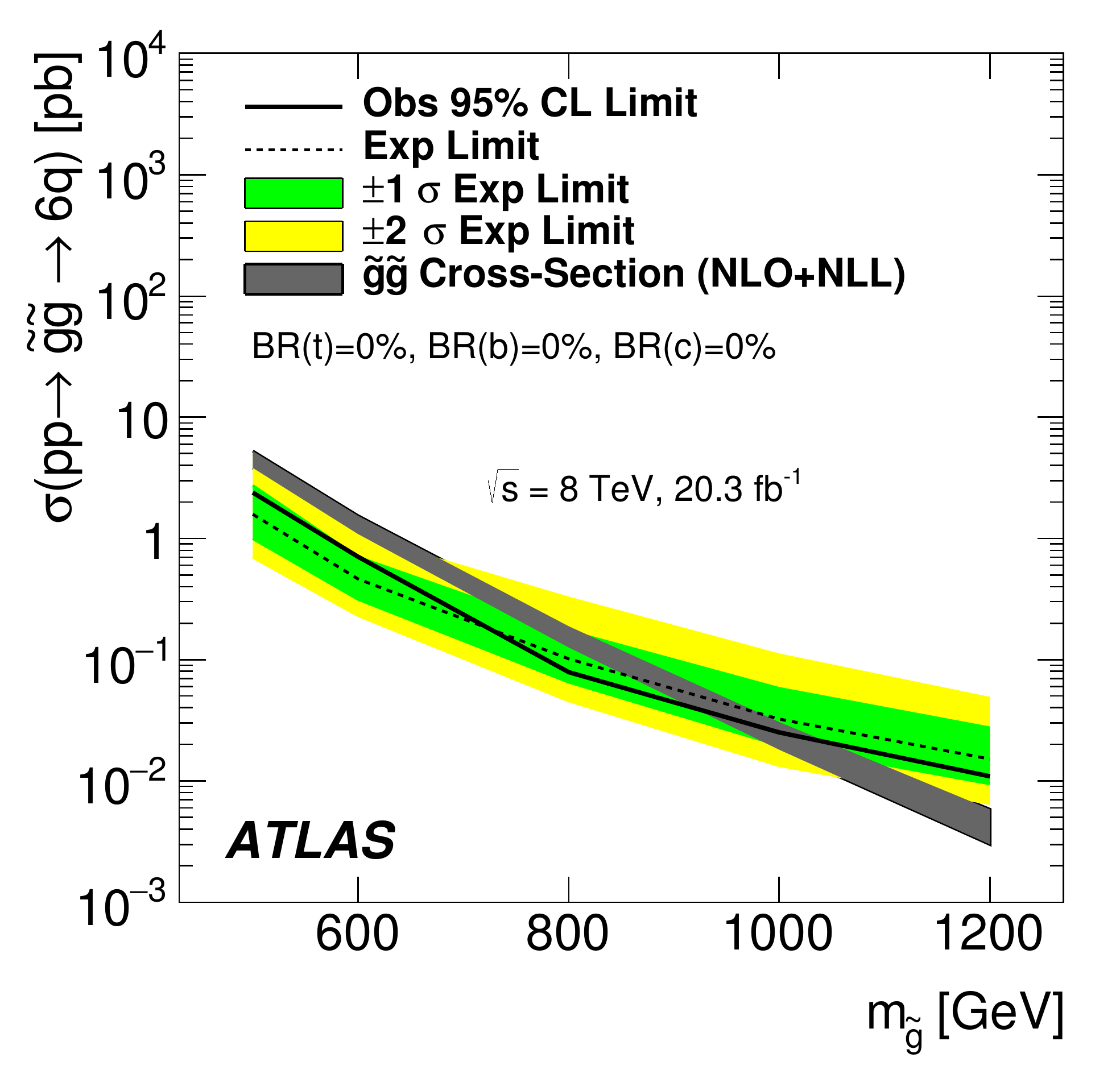}
\caption{BR($b$)=0\% - Observed}
\end{subfigure}
\begin{subfigure} {0.5\textwidth}
\centering
    \includegraphics[width=0.9\textwidth]{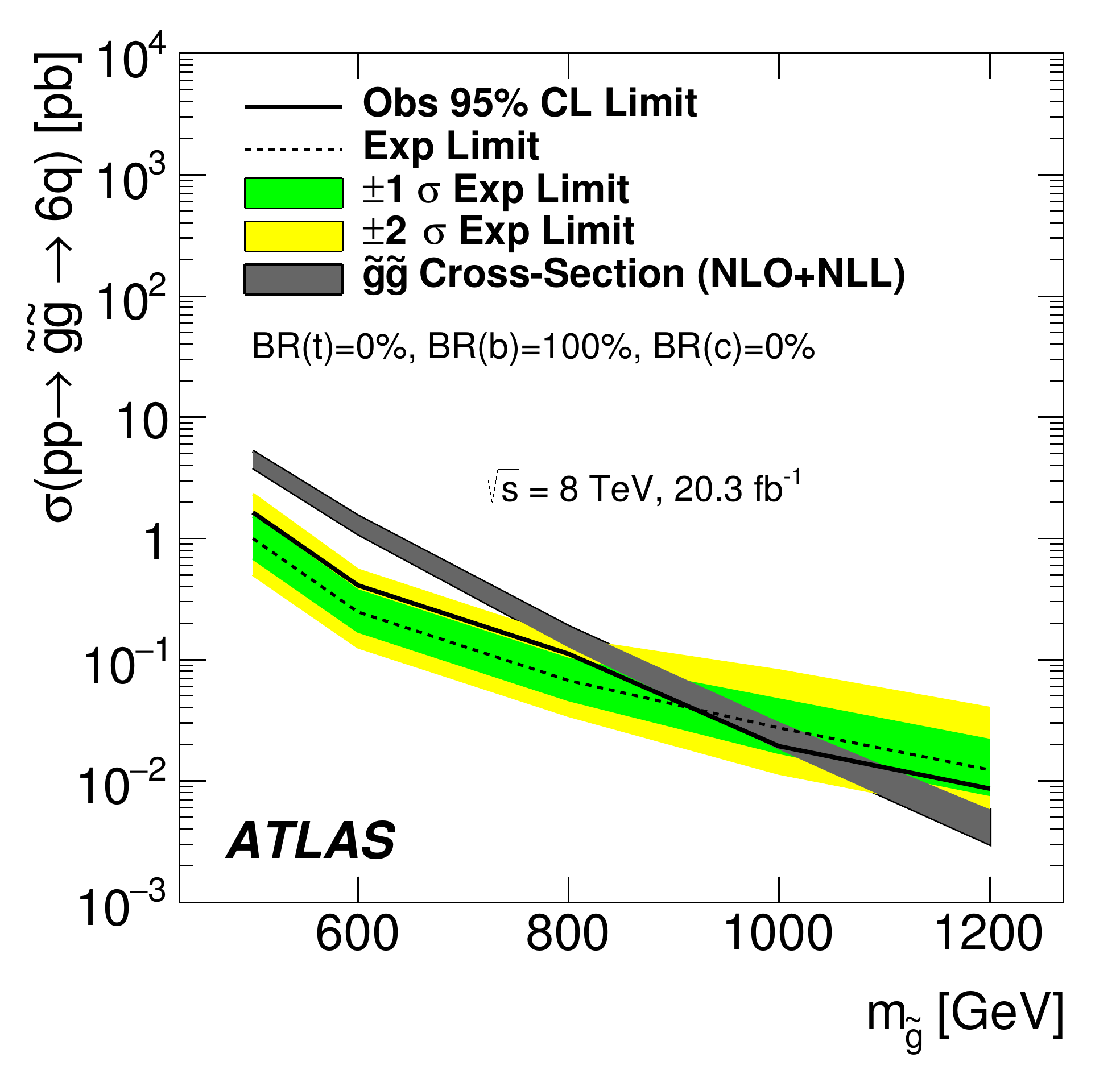}
\caption{BR($b$)=100\% - Observed}
\end{subfigure}
\caption{Observed mass exclusions at the 95\% CL for BR($b$)=0\%  (left) and  BR($b$)=100\% (right).  (from~\cite{Aad:2015lea}).}
\label{fig:mass_BR-bjets-multij-A}
\end{figure}

More generally, excluded masses as a function of the branching ratios of the decays are presented in \figref{fig:massSummary6q_6-multij-A}  where each bin shows the maximum gluino mass that is excluded for the given decay mode. 
It is illustrative to recognize the observed mass limit from~\figref{fig:mass_BR-bjets-multij-A} (a) can also be found in the lower left corner of~\figref{fig:massSummary6q_6-multij-A}.

\begin{figure}[htbp]
\begin{subfigure} {0.5\textwidth}
\centering
    \includegraphics[width=0.9\textwidth]{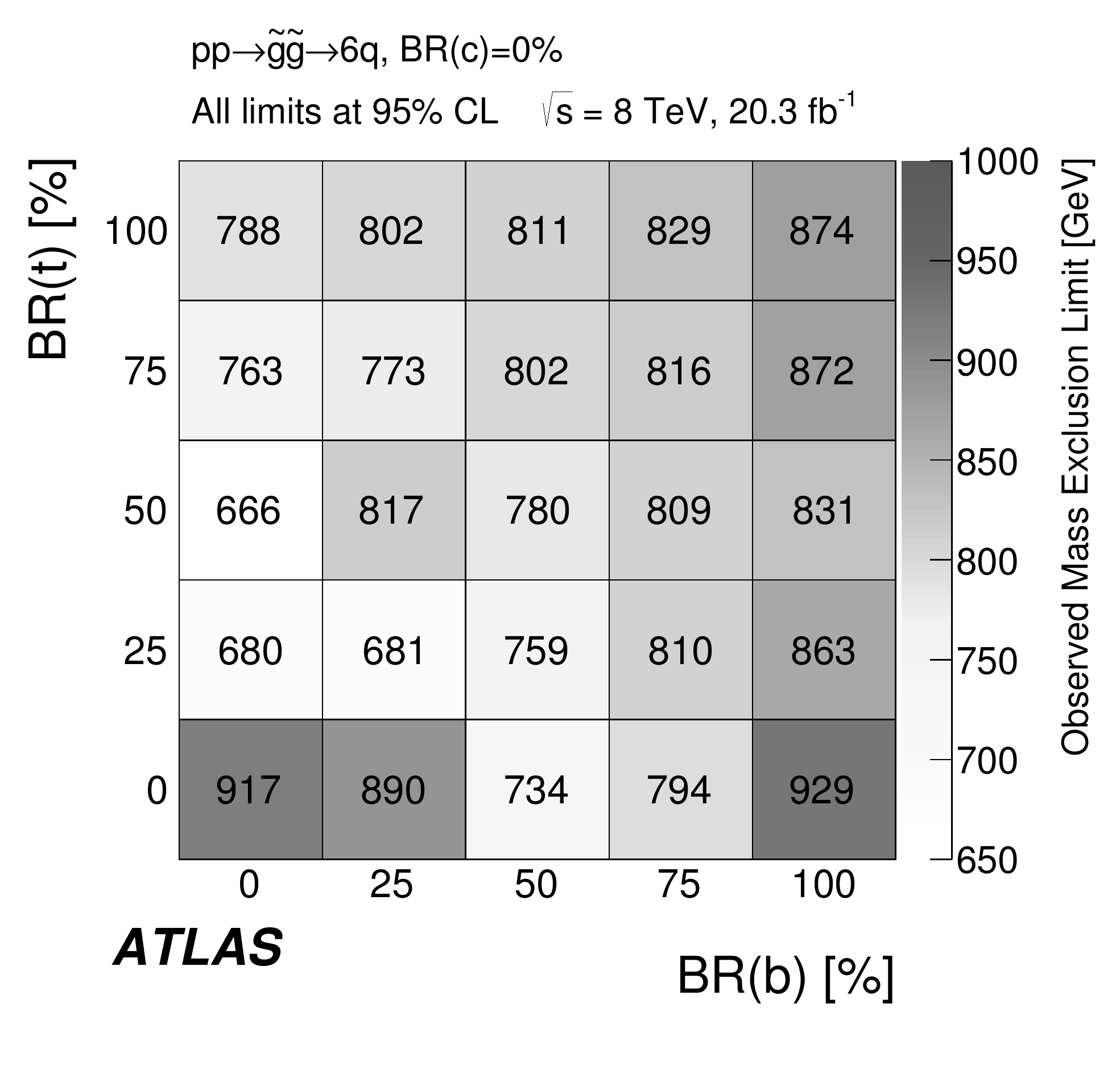}
\caption{BR($c$)=0\% - Observed}
\end{subfigure}
\begin{subfigure} {0.5\textwidth}
\centering
    \includegraphics[width=0.9\textwidth]{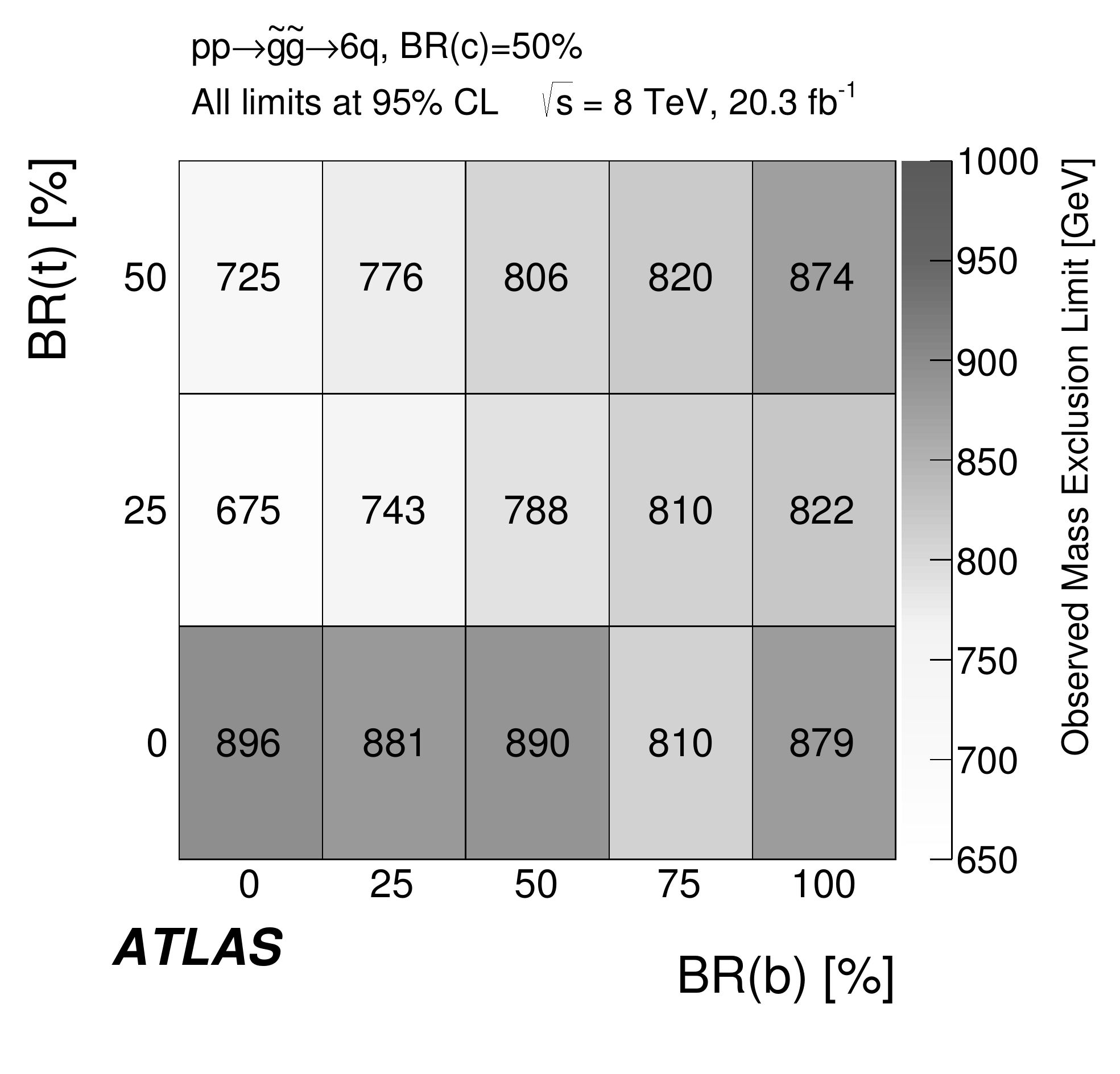}
\caption{BR($c$)=50\% - Observed}
\end{subfigure}
\caption{Observed mass exclusions at the 95\% CL in the BR($t$) vs. BR($b$) space for BR($c$)=0\% (left) and  BR($c$)=50\% (right). Each point in this space is individually optimized and fit. Masses below these values are excluded in the six-quark model. Bin centers correspond to evaluated models  (from~\cite{Aad:2015lea}).}
\label{fig:massSummary6q_6-multij-A}
\end{figure}

The interpretations of the results of the jet-counting and total-jet-mass analyses are displayed together in \figref{fig:overlaidcontours_10-multij-A} for the ten-quark model. This figure allows for the direct comparison of the results of the various analyses. 
Without $b$-tagging requirements, the jet-counting analysis sets slightly lower expected limits than the total-jet-mass analysis. 
With $b$-tagging requirements, the limits are stronger for the jet-counting analysis. 
The observed limits from the total-jet-mass analysis and jet-counting analysis with $b$-tagging requirements
 are also comparable.
\begin{figure}[!htb]
\centering
\includegraphics[width=0.6\textwidth]{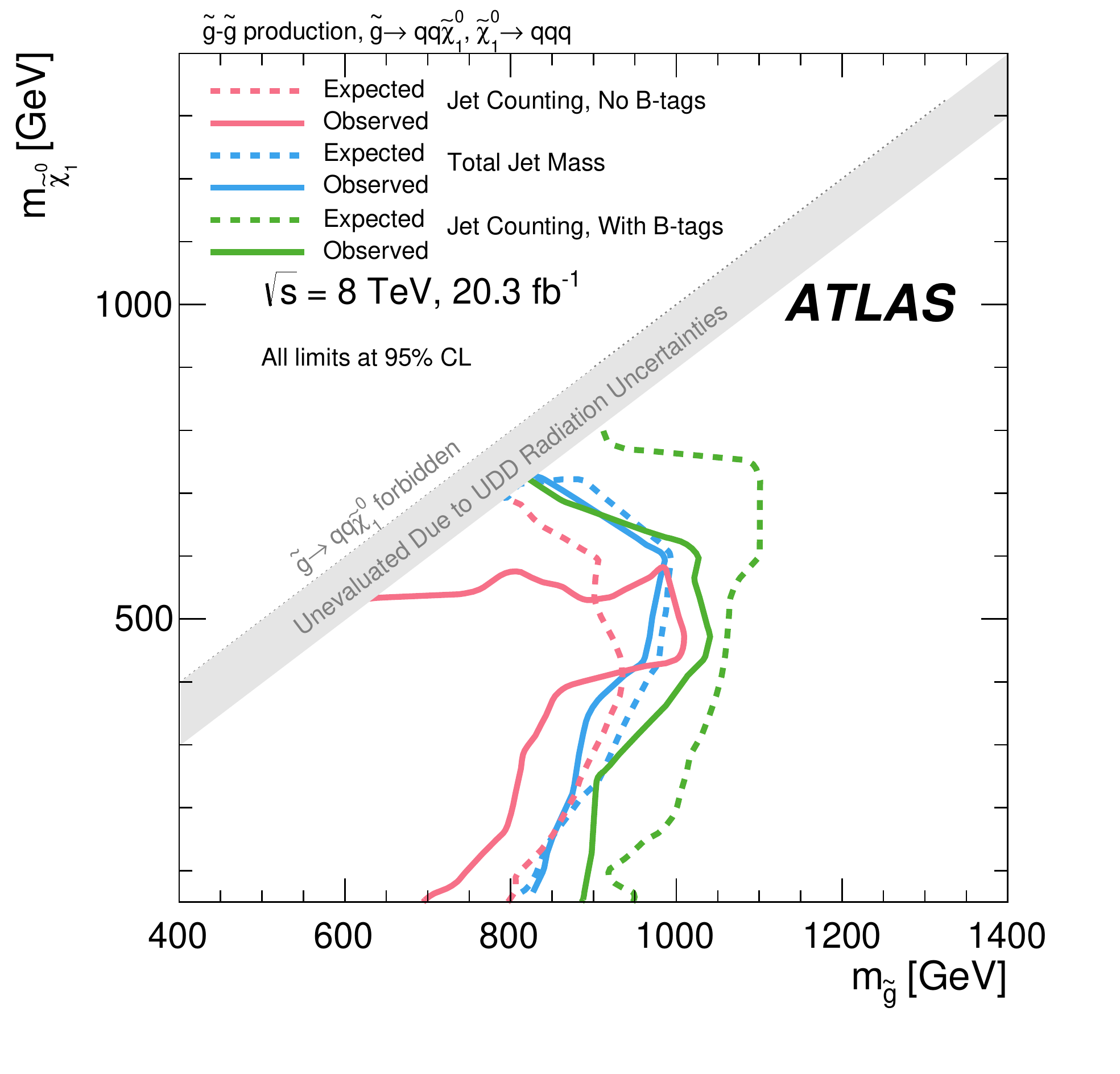}
\caption{Expected and observed exclusion limits in the  ($m_{\gluino}-m_{\ninoone}$) plane for the ten-quark model for the jet-counting analysis (with and without $b$-tagged jets) and the total-jet-mass analysis  (from~\cite{Aad:2015lea}).}
\label{fig:overlaidcontours_10-multij-A}
\end{figure}

Exclusion limits at the 95\% CL are set extending up to $m_{\gluino}=917$~GeV in the case of pair-produced gluino decays to six light quarks and up to $m_{\gluino}=1$~TeV in the case of cascade decays to ten quarks for moderate $m_{\gluino}-m_{\ninoone}$ mass splittings.

It is interesting to note that strong model-independent limits have been obtained in~\cite{Aad:2015lea}. 
Within the jet-counting method, the 95\% CL upper limits obtained on the
(observed) \emph{visible signal cross-section} vary from 0.2~fb to 2.6~fb, 
depending on the requirements on jet \pt\ and number of $b$-jets for different SRs. 
Notably the strongest limit of 0.2~fb has been derived in SR requiring
at least seven jets, with $\pt$ above 180~GeV and one $b$-tagged jet.

\subsection{\gluino\ production with leptonic final states at ATLAS}
\label{sec:UDD_SS3L_models_ATLAS}
In the gluino-mediated top squark $\to bs$ (RPV) model investigated in~\cite{Aad:2014pda}, top squarks are assumed to decay with the \UDD coupling
$\lambda''_{323}=1$.
The final state is therefore  $\gluino\gluino\to bbbb~ss~WW$,
characterized by the presence of four $b$-quarks and only moderate missing transverse momentum. 

Results are interpreted in the parameter space of the gluino and top squark masses (see figure~\ref{fig:UDD_323_SS3L}).
Gluino masses below 850~\gev\  are excluded at 95\% CL, almost independently of the stop mass.
The sensitivity is dominated by SR3b. 
The SR3b signal region is sensitive to various models with same-sign or
$\geq$ 3 leptons and $\geq$ 3 $b$-quarks.
This is also demonstrated in the gluino-mediated top squark $\to bs$ (RPV) model, 
where $m_{\gluino}<850$~\gev\ is excluded by SR3b alone in the absence of a large \met\ signature.

It is important to mention that for the same simplified model, a similar bound of $m_{\gluino}>900$~\gev\ has been obtained in the ATLAS search for (7 -- 10) jets and \met~\cite{Aad:2013wta}.
The latter exclusion limit tends to be extended for relatively light or heavy stops. 

\begin{figure}[htbp]
\centering
\includegraphics[width=0.6\textwidth]{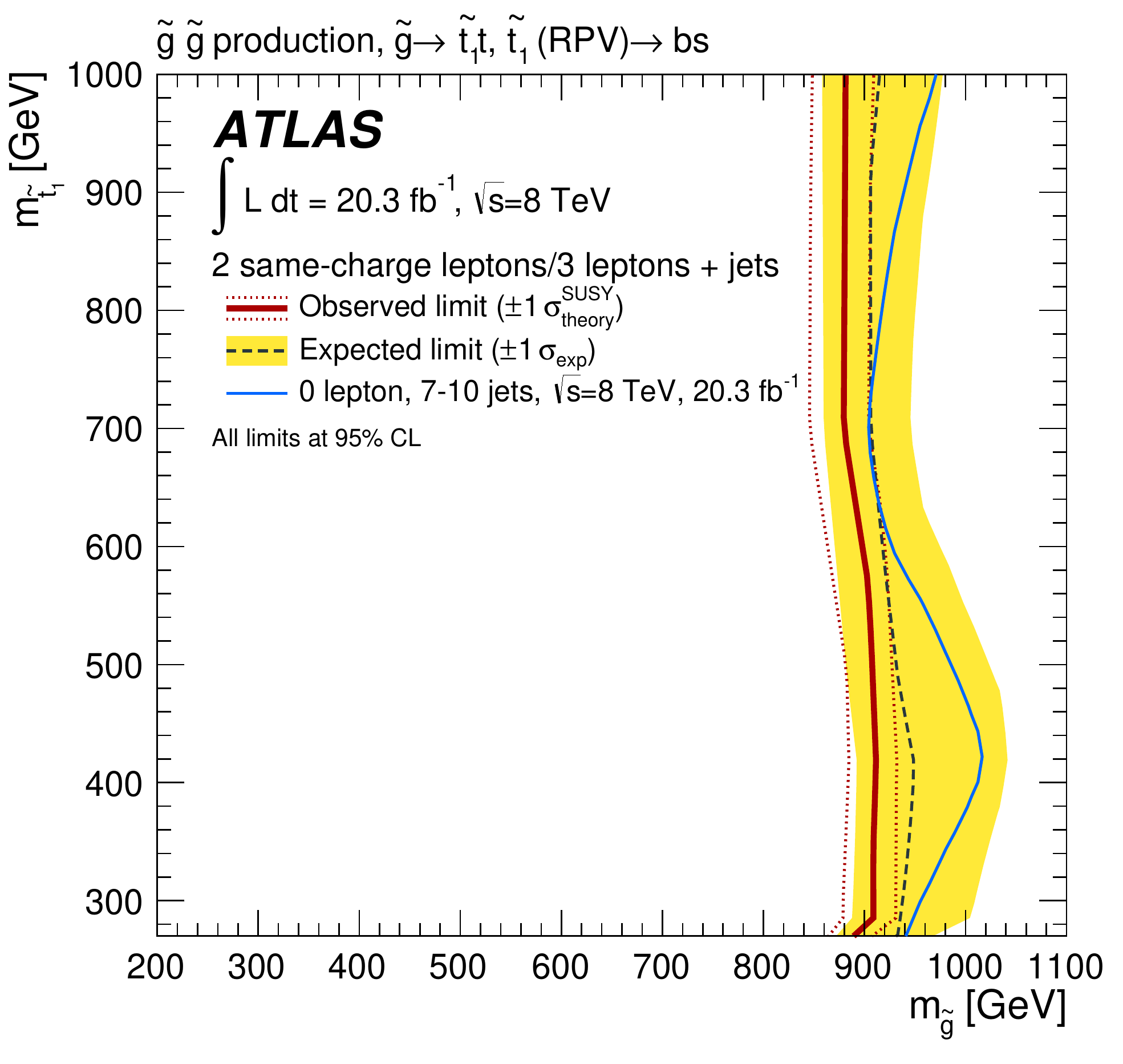}
   \caption{Observed and expected exclusion limits on gluino-mediated top squark production, 
obtained with 20.3\,fb$^{-1}$ of $pp$ collisions at $\sqrt{s}$ = 8~\TeV, for the
top squark decay modes via $\lamdp[323]$ (from~\cite{Aad:2014pda}).
      }
   \label{fig:UDD_323_SS3L}
\end{figure}
As a model-independent limit from SR3b, the limit on the visible cross-section $\sigma_{\textrm{vis}}=0.19$~fb has been obtained at the 95\% CL.
It is interesting to note that SR3b is also the most sensitive signal region constraining \bRPV mSUGRA.

\subsection{\gluino\ production with multi-jets at CMS}
\label{sec:UDDmodels_CMS_multij_C}
The signal is modeled in~\cite{Chatrchyan:2013gia} with pair-produced gluinos where each gluino decays to three quarks through \UDD-type couplings. 
Two different scenarios, an inclusive search and also a heavy-flavor search,
are considered in that analysis.
For the first case,
the coupling $\lambda''_{\textrm{112}}$
is set to a non-zero value, giving a branching fraction of 100\% for the gluino decay to
three light-flavor quarks.
The second case, represented
by $\lambda''_{\textrm{113}}$ or $\lambda''_{\textrm{223}}$, investigates
gluino decays to one $b$ quark and two light-flavor quarks.
In these simplified models, all superpartners
except the gluino are taken to be decoupled, the natural
width of the gluino resonance is assumed to be much smaller than the
mass resolution of the detector,
and no intermediate particles are produced in
the gluino decay.

As is also illustrated in~\figref{fig:UDD_112_113-223_multi-jets_C}, several constraints on \gluino-masses have been derived in~\cite{Chatrchyan:2013gia}.
The production of gluinos undergoing RPV decays into light-flavor jets is
excluded at 95\% CL for gluino masses below 650~\GeV, with a less conservative exclusion of 670~\GeV\ based 
upon the theory value at the central scale. The respective expected limits are
755 and 795~\GeV.
Gluinos whose decay includes a heavy-flavour jet are
excluded for masses between 200 and 835~\GeV, with the less conservative exclusion up to 855~\GeV\ from the central theoretical value. The respective expected limits are 825 and 860~\GeV.
In the heavy-flavor search the limits extend to higher masses because of the reduction of the background.

\begin{figure}[htbp]
\begin{subfigure}{0.5\textwidth}
\centering
\includegraphics[width=0.9\textwidth]{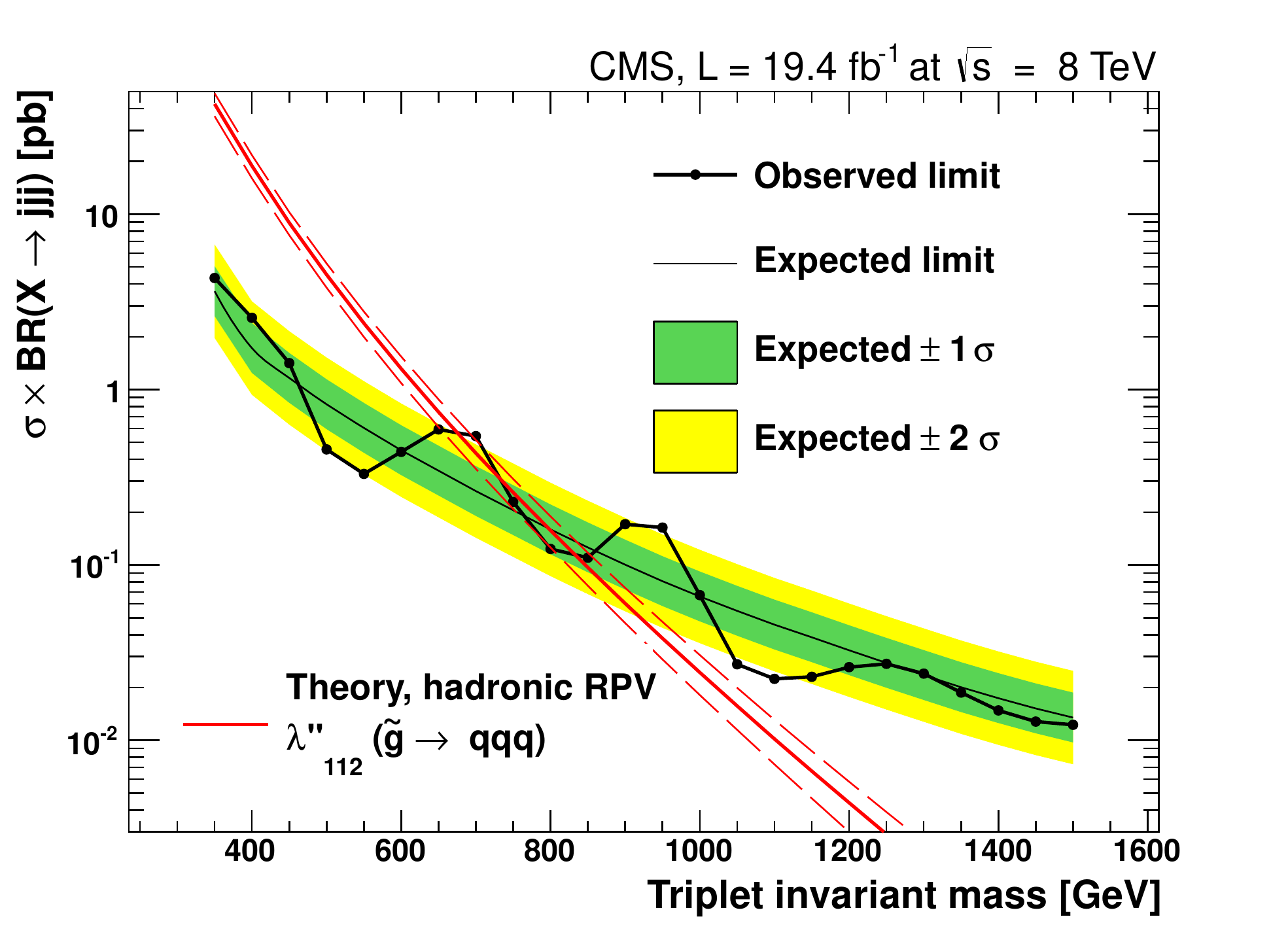}
\caption{~\lamdp[112]}
\end{subfigure}
\begin{subfigure}{0.5\textwidth}
\centering
\includegraphics[width=0.9\textwidth]{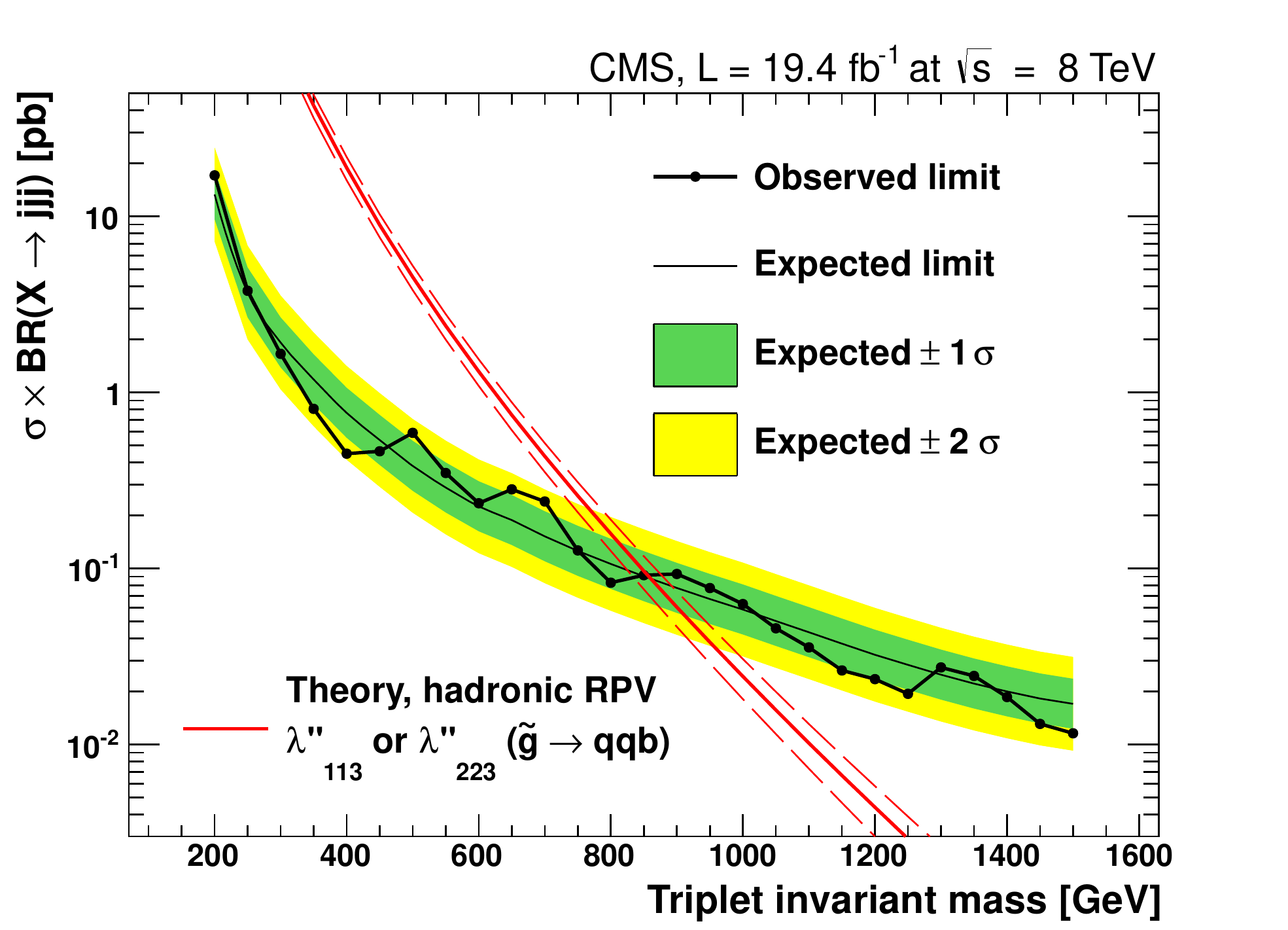}
\caption{~\lamdp[113] or ~\lamdp[223]}
\end{subfigure}
   \caption{Observed and expected 95\% CL cross-section limits as a function of mass
         for the inclusive (left) and heavy-flavor searches (right).
	  The limits for the heavy-flavor
	 search cover two mass ranges, one for low-mass gluinos ranging from 200 to 600~\GeV,
	 and one for high-mass gluinos covering the complementary mass range up
	 to 1500~\GeV\ (from~\cite{Chatrchyan:2013gia}). 
      }
   \label{fig:UDD_112_113-223_multi-jets_C}
\end{figure}

\subsection{\gluino\ production with same-sign leptons at CMS}
\label{sec:UDD_SS_models_CMS}
In this analysis~\cite{Chatrchyan:2013fea}, a simplified model based on gluino pair production followed
by the decay of each gluino to three quarks is considered.
It is interesting to note that the analogous model is also taken into account in~\cite{Aad:2014pda}, as mentioned above.
Moreover, \UDD-like decays can in principle be motivated also from the SUSY model
with minimal flavour violation~\cite{Csaki:2011ge}.
In~\cite{Chatrchyan:2013fea}, the focus is on the decay mode
$\gluino\to t b s$. 
Due to its Majorana nature, the corresponding anti-particles emerge with equal probability in the decay of \gluino. 
Such decays lead to
same-sign W-boson pairs in the final state in 50\% of the cases. 

The signal process is illustrated in~\figref{fig:diag_gluino_stop_SS_C}.
In comparison to the decays
$\gluino\to t s d$, yielding also same-sign $W$-boson pairs,
the mode $tbs$ is investigated. Due to two extra $b$ quarks in the final state a higher signal selection
efficiency can finally be obtained. 
The key parameter of the model is $m_{\gluino}$ determining the production cross-section and the final state
kinematics. The dedicated search region RPV2 with the high-$\pt$ lepton selection
is used to place an upper limit on the production cross-section.

\begin{figure}[htbp]
\centering
\includegraphics[width=0.4\textwidth]{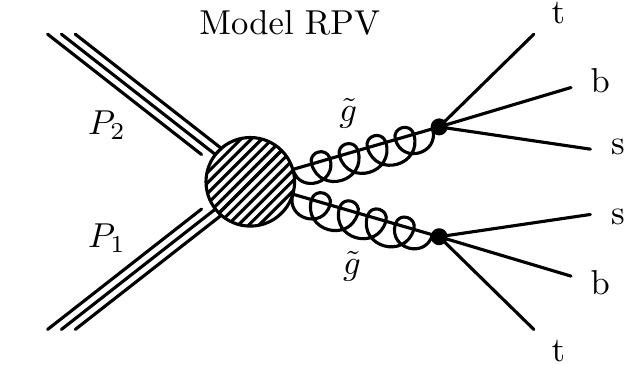}
   \caption{Signal process for $\gluino\to t b s$ assuming gluino pair-production and $\lamdp[323]$ coupling.
      }
   \label{fig:diag_gluino_stop_SS_C}
\end{figure}

The result is shown in Fig.~\ref{fig:UDD_323_SS_C}. 
In this scenario, the gluino mass is probed up to approximately 900~\GeV.
A similar exclusion limit from the corresponding ATLAS search has been obtained, as discussed in Section~\ref{sec:UDD_SS3L_models_ATLAS}.  

\label{sec:UDD_SS_results_CMS}
\begin{figure}[htbp]
\centering
\includegraphics[width=0.6\textwidth]{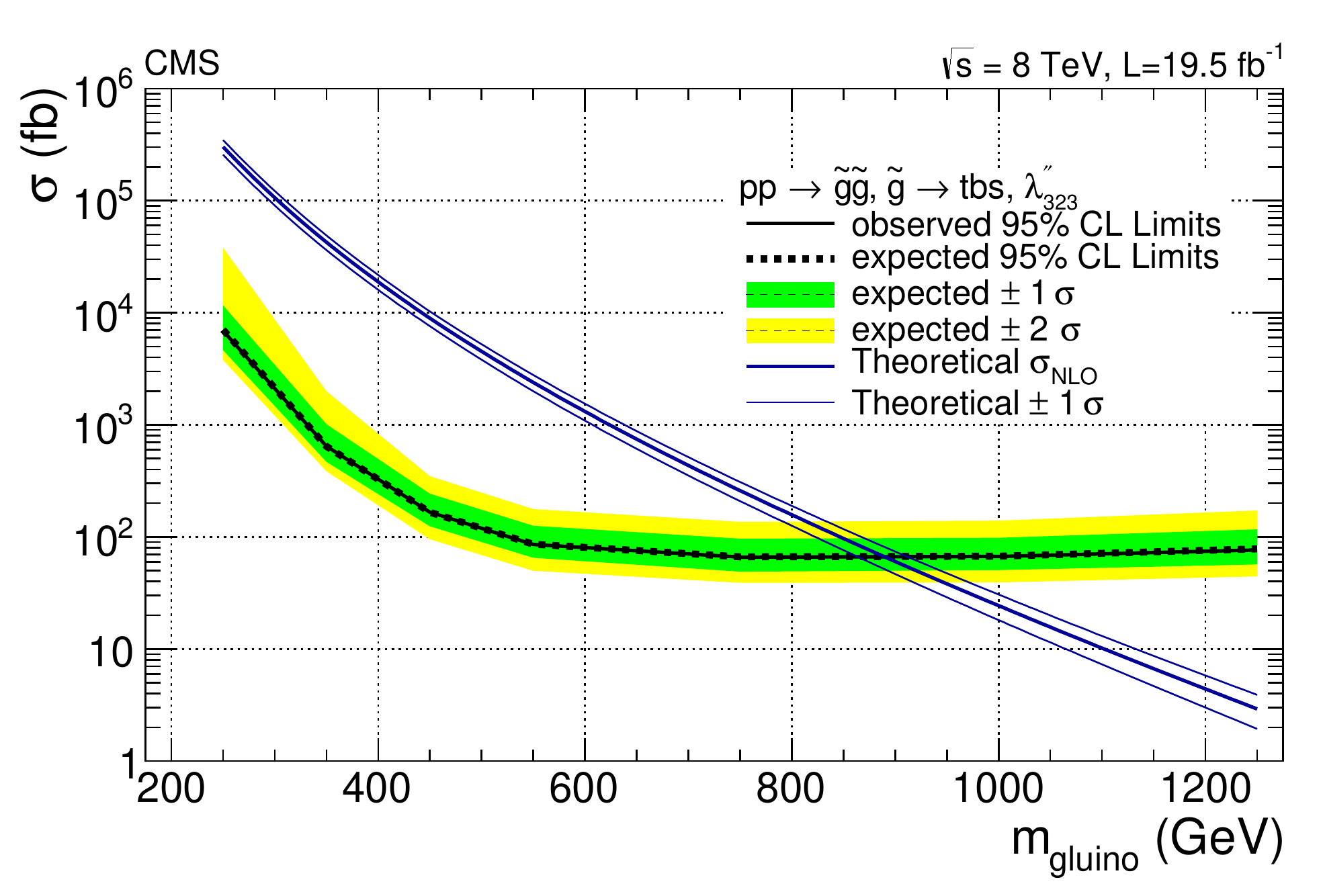}
   \caption{95\% CL upper limit on the gluino production cross-section for a RPV simplified model,
$p p \to \gluino \gluino, \gluino \to t b s$ (from~\cite{Chatrchyan:2013fea}).
      }
   \label{fig:UDD_323_SS_C}
\end{figure}

\subsection{\stopone\ production with jet pairs at CMS}
\label{sec:UDDmodels_CMS_pairj_C}
The analysis~\cite{Khachatryan:2014lpa} has been optimized by studying two simplified models with stop pair production:
First, the coupling $\lamdp[312]$ is assumed leading to two light-flavor jets in the decay of each \stopone. 

Considering $\lamdp[323]$ non-zero in the second simplified model, 
one $b$-jet and
one light-flavor jet are generated per $\stopone$.
In both of the above cases, the branching ratio of the top squark decay to two jets
is set to 100\% and all superpartners except the top squarks
are taken to be decoupled, 
so that no intermediate particles are produced in the top squark decay.

Figure~\ref{fig:UDD_312_323_jetpair_C} shows the observed and expected
95\% CL upper limits obtained in~\cite{Khachatryan:2014lpa}
 based on results from the low-mass and high-mass SRs, respectively.
In that case the top squark mass corresponds to $m_{\text{av}}$.
The vertical dashed blue line at a top squark mass of 300~\GeV\
indicates the transition from the low- to the high-mass limits,
and at this mass point the limits are shown for both analyses.
The production of top squarks decaying via $\lamdp[312]$ into light-flavor
jets is excluded at 95\% CL for top squark masses from 200 to 350~\GeV.
Stops decaying via \lamdp[323] coupling, thus leading to a heavy-flavor jet, are excluded
for masses between 200 and 385~\GeV.

\begin{figure}[htbp]
\begin{subfigure}{0.5\textwidth}
\centering
\includegraphics[width=0.9\textwidth]{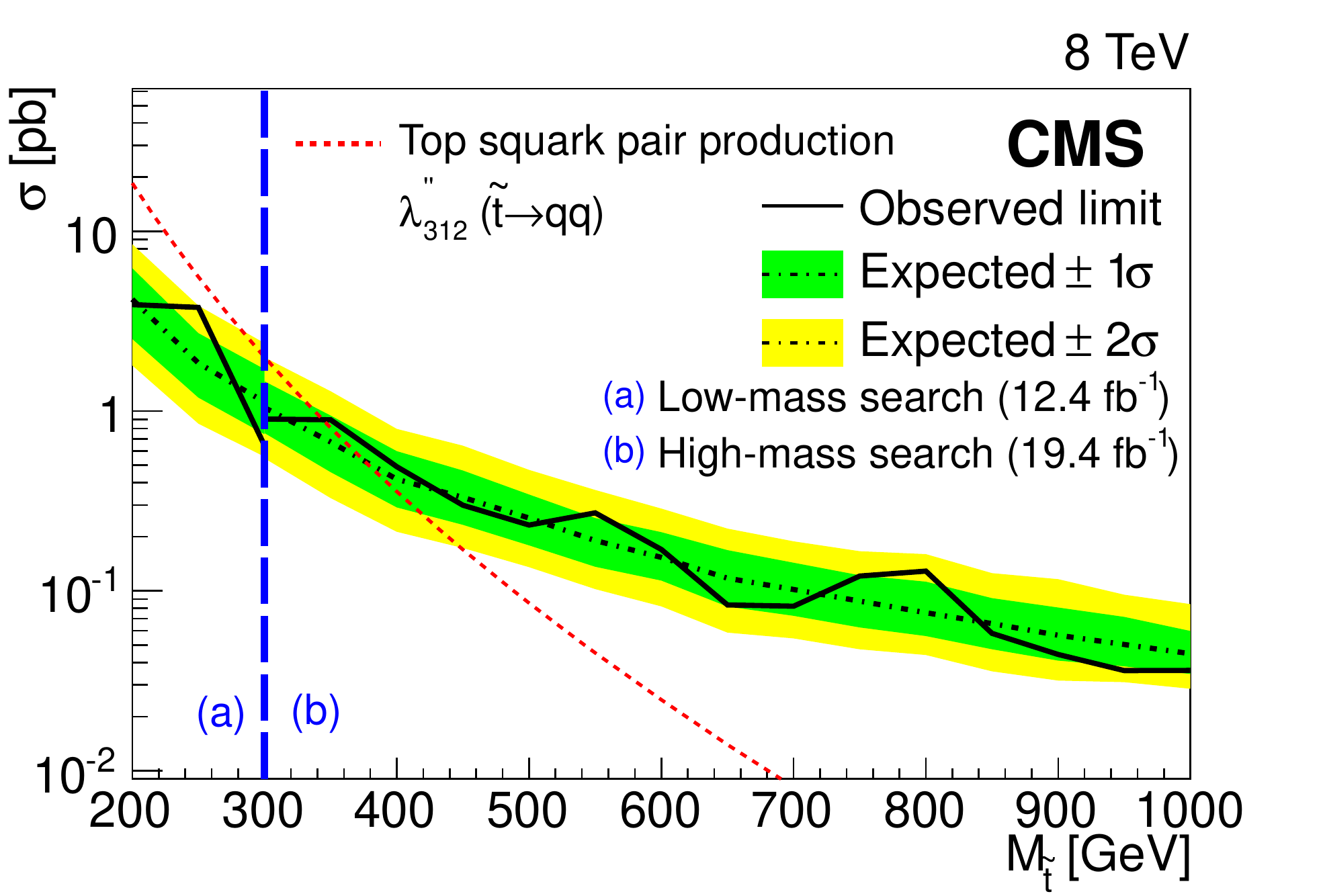}
\caption{~\lamdp[312]}
\end{subfigure}
\begin{subfigure}{0.5\textwidth}
\centering
\includegraphics[width=0.9\textwidth]{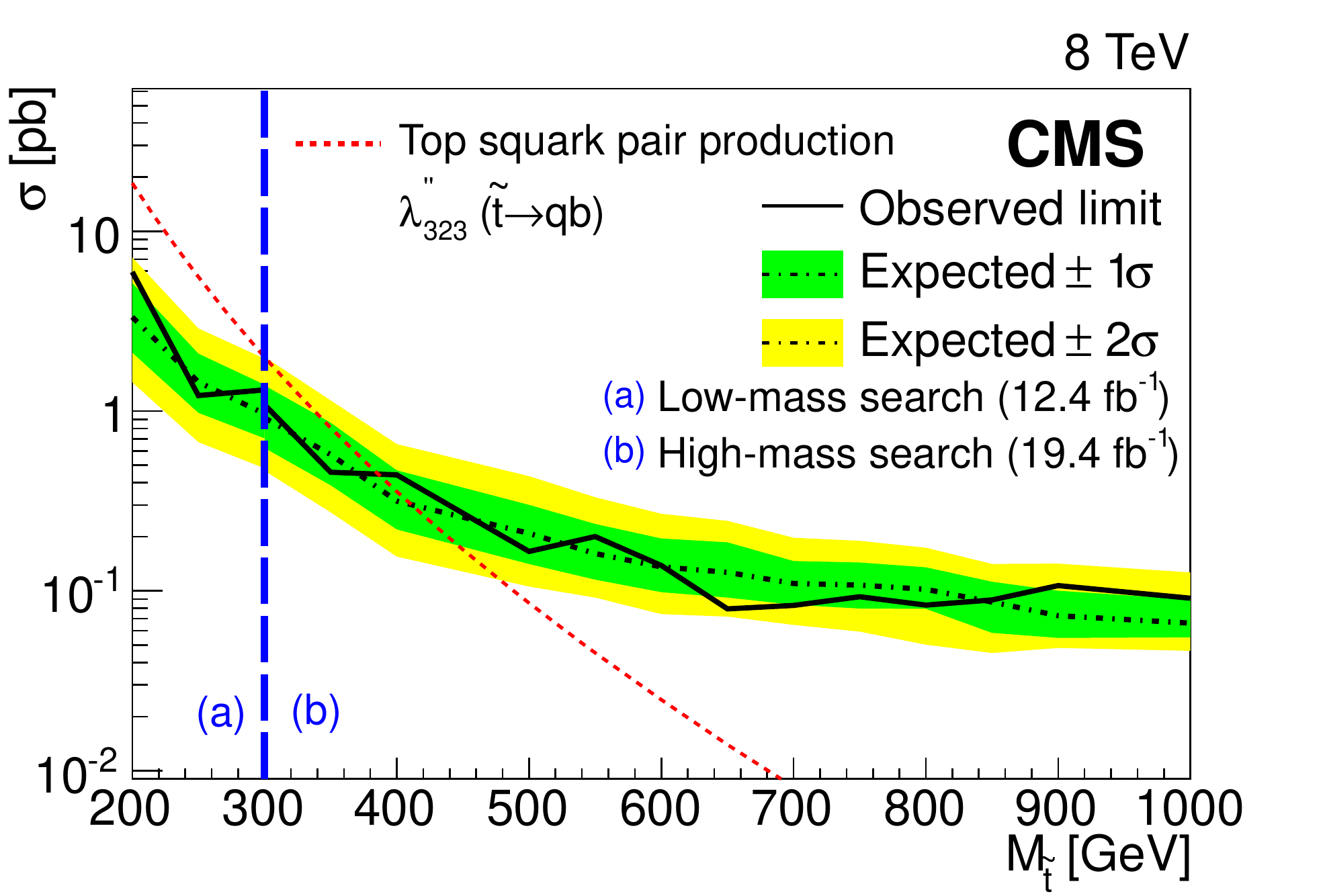}
\caption{~\lamdp[323]}
\end{subfigure}
   \caption{Observed and expected 95\% CL cross-section limits as a function of top squark mass
         for the inclusive (left) and heavy-flavor (right) searches for \RP-violating stop decays
         (from~\cite{Khachatryan:2014lpa}).
      }
   \label{fig:UDD_312_323_jetpair_C}
\end{figure}

\section{Conclusions}
\label{sec:conclusions}

Results of searches for signatures of prompt \RP violation at 8~TeV at LHC experiments have probed RPV SUSY at the highest collider energies so far. 
No significant deviations have been found in the corresponding ATLAS and CMS analyses, implying strong constraints on superpartner masses and/or RPV couplings.
A common assumption for many interpretations are \ninoone\ LSPs with varying assumptions on dominant \RP-violating couplings and in particular NLSP types.
Using simplified models with RPC production of NLSPs and subsequent decays via LSPs and \LLE interactions, the following approximate upper limits on superpartner masses have been obtained:
\begin{itemize}
\item Gluino masses $m(\gluino)>950$~\gev.
\item Light stop masses $m(\stopone)>820$~\gev.
\item Wino-like chargino masses $m(\chinoonepm)>450$~\gev.
\item Charged slepton masses $m(\tilde{l})>240$~\gev.
\item Sneutrino masses $m(\tilde{\nu})>400$~\gev.
\end{itemize}
Resonance searches have mainly focused on analyzing heavy narrow resonances of tau-sneutrinos, excluding masses up to 2.0~\tev, thus extending previous limits from Tevatron significantly.
Limits based on dominant \LQD couplings have been investigated in models with stop-pair production,
constraining stop masses up to 1~TeV.
Relaxing the assumption of dominance of a single \RP-violating coupling has, for example, been investigated in the \UDD multi-jet analysis by ATLAS: Variation of corresponding branching ratios to different heavy quarks has lead to upper limits of gluino  masses within the range 666~\gev$< m(\gluino) < $929~\gev. 
In contrast to trilinear RPV models, searches for bilinear RPV have assumed mSUGRA boundary conditions, yielding the first collider-based observed limits for \bRPV models: 
Requiring mSUGRA parameters consistent with the observed mass of the Higgs boson, limits from \bRPV searches exclude 
 gluino masses in that model around 1.3~\tev.    

The strongest model-independent limits for observed visible
cross-sections have been derived at approximately 0.2~fb.
It is interesting to note that such a strong constraint has been obtained 
in the following searches:
\begin{itemize}
\item Multi-leptons in SRs requiring at least three light leptons.
\item Same-sign or three leptons in combination with at least three $b$-jets.
\item Two hadronically decaying taus in conjunction with jets and \met.
\item Seven jets, with $\pt$ above 180~GeV and one $b$-tagged jet.
\end{itemize}
Summarizing the relevant signal regions defined for these searches at the ATLAS and CMS experiments
also facilitates identifying possible new targets for future analysis optimization.
This should include improved reconstruction of highly collimated objects with low \met, relevant in scenarios predicting strongly boosted final states. 

Obviously the whole parameter space of RPV SUSY has not been covered in LHC searches.
Various options in particular for investigating \LQD couplings remain and should be subject of systematic studies.
Indeed most of the limits for prompt RPV from Run I have been obtained assuming either \LLE or \UDD couplings in simplified models.
Therefore it would be interesting not only to vary the types of RPV couplings, however also
consider approaches for studying complete SUSY mass spectra with different \RP-violating couplings. 
As an example, extending pMSSM models with RPV decays would lead to significantly different final states topologies in comparison to the RPC-based pMSSM models analyzed frequently. 
Also considering alternative options for the nature of both the NLSP and the LSP would modify some of the model-dependent  results mentioned before.
As an example, the assumption of a stau LSP has only been investigated in the analysis of 7~\tev\ data implying different final states with respect to \ninoone\ LSPs. 
It would also be interesting to search for various types of heavy sparticle RPV resonances using the increased future energies at the LHC. 

Since the largest cross-sections are predicted for supersymmetric strong production processes for LHC Run II,
signatures from gluino and/or squark production typically offer high potential for future RPV searches.
Increasing the luminosity will also enhance the sensitivity for searches focusing on electroweak production processes. 
Ultimately, the results for RPV SUSY in Run II can become crucial for the question of supersymmetry at the weak scale.

\section*{Acknowledgments}
The author is grateful to J. Boyd, M. Flowerdew, T. Lari, and V. Mitsou for
their comments on this paper. This work was supported by the German BMBF within
the research network FSP-101 "Physics on the TeV Scale with ATLAS at the LHC"
and by the German Helmholtz Alliance "Physics at the Terascale."

\bibliographystyle{useBibStyleWoTitle}
\bibliography{RPV_review_LHC.bib}


\end{document}